\documentstyle[epsf]{mn}

\newcommand{\bcen}{\begin{center}}
\newcommand{\ecen}{\end{center}}
\newcommand{\bfig}{\begin{figure}}
\newcommand{\efig}{\end{figure}}

\newcommand{\scrh}{{\cal H}}

\newcommand{\half}{\frac{1}{2}}

\newcommand{\kms}{{\rm\,km\,s^{-1}}}

\newcommand{\mpc}{{\rm\,Mpc}}
\newcommand{\msun}{{\rm\,M_\odot}}

\def\pc{{\rm\,pc}}

\newcommand{\yr}{{\rm\,yr}}

\title[Stellar Dynamics in Galactic Nuclei]
{Stellar Dynamics around Black Holes in Galactic Nuclei}
\author[S.~Sridhar and J.~Touma]
        {S.~Sridhar\thanks{E-mail: sridhar@iucaa.ernet.in} and
         J.~Touma\thanks{E-mail: touma@harlan.as.utexas.edu}\\
        $^*$~Inter-University Centre for Astronomy and Astrophysics, 
        Ganeshkhind, Pune 411 007, INDIA\\
        $^{\dag}$~University of Texas, McDonald Observatory, 
        RLM 15.308, Austin, Texas, 78712}

\begin{document}
\label{firstpage}
\maketitle

\begin{abstract} 
We classify orbits of stars that are bound to central black holes in
galactic nuclei. The stars move under the combined gravitational
influences of the black hole and the central star cluster. Within the
sphere of influence of the black hole, the orbital periods of the
stars are much shorter than the periods of precession. We average over
the orbital motion and end up with a simpler problem and an extra
integral of motion: the product of the black hole mass and the
semimajor axis of the orbit. {\em Thus the black hole enforces some
degree of regularity in its neighborhood}. Well within the sphere of
influence, (i) planar, as well as three dimensional, axisymmetric
configurations--both of which could be lopsided--are integrable, (ii)
fully three dimensional clusters with no spatial symmetry whatsover
must have semi--regular dynamics with two integrals of motion. Similar
considerations apply to stellar orbits when the black hole grows
adiabatically.  We introduce a family of planar, non--axisymmetric
potential perturbations, and study the orbital structure for the
harmonic case in some detail.  In the centered potentials there are
essentially two main families of orbits: the familiar loops and
lenses, which were discussed in Sridhar and Touma (1997, MNRAS, 287,
L1-L4). We study the effect of lopsidedness, and identify 
a family of loop orbits, whose orientation reinforces the lopsidedness,
an encouraging sign for the construction of self--consistent models of
eccentric, discs around black holes, such as in M31 and NGC 4486B.
\end{abstract}
\begin{keywords}
galaxies:  kinematics and dynamics
\end{keywords}

\section{Introduction}

Following up on a decade of work with ground--based telescopes (cf.
Kormendy~1982, 1987 and Lauer~1983, 1985), observations with the
Hubble Space Telescope (HST) have clarified, and extended our
knowledge of the centres of galaxies (cf. Lauer et. al 1995, Gebhardt
et al. 1996).  The brightness profiles are generically cuspy, and many
correlations have been drawn between cusp steepness, isophote shapes,
luminosities, kinematics, and other nuclear and global properties of
galaxies. Most, if not all, galaxies might have central black holes
(BH), massive enough to power active nuclei when there is an adequate
supply of fuel (Rees~1990).  About a dozen galaxies have been
shortlisted as candidates for possessing central BHs (c.f. Kormendy
and Richstone~1995), and the case appears strong for our Galaxy,
M31, M32, and NGC 3115.  BH detection based on spectroscopy of stellar
light requires careful modeling of the distribution and motions of
stars in the central regions.  Set against this background of
progress, our understanding of the dynamics of nuclear star clusters
is quite poor. Gerhard and Binney (1985) argued that cusps and BHs
will destroy box orbits, which are
then replaced by chaotic orbits. The
chaotic orbits being rounder,  it is difficult to construct strongly
non-axisymmetric equilibria (c.f. Merritt~1997, 1998 for a flavour of 
recent work in this area). Sridhar
and Touma (1997) constructed, cuspy, non-axisymmetric, scale-free
discs whose potentials are of St\"ackel form in parabolic
coordinates. The dynamics in these potentials is fully integrable. A
BH could be added at the centre without affecting the integrability of
motion. Discs without a BH support only ``banana'' orbits. The BH stabilizes a
box-like family of orbits, the lenses. These models, as they questioned
the implicit assumption that cusps and BHs imply chaos, perked our
interest in further investigating dynamics very close to the BH. Here,
we explore another route to more regular dynamics in galactic nuclei.

The orbits of stars in galactic nuclei are controlled by the combined
gravitational forces of the central BH, and the self gravity of the
cluster.  Very close to the BH, the orbits are nearly
Keplerian---stars bound to the BH move on nearly elliptical orbits,
whereas very energetic ones follow hyperbolic trajectories (we ignore
relativistic effects, so our discussion is applicable only to
distances beyond several Schwarzschild radii from the BH). If the
velocity dispersion in the cluster is $\sigma$, we might say that a BH
of mass $M$ has a strong influence on stellar orbits inside a sphere
of radius, $r_h\sim GM/\sigma^2\,$. For fiducial values, $M\sim
10^8\msun\,$, $\sigma\sim 100\kms\,$, we obtain $r_h\sim 50\pc\,$, a
spatial scale that HST with a resolution of $0.''1$ can easily resolve
much farther than the Virgo cluster.\footnote{It should be noted that $\sigma$
itself is often not independent of distance from the BH. In this case, $\sigma$
should be taken as the dispersion at $r=r_h$, which results in an 
equation wherein $r_h$ is the unknown quantity to be solved for. 
Perhaps a better estimate of the
sphere of influence of the BH is the radius at which the mass of the
cluster equals the mass of the BH.} Keplerian ellipses do not precess,
and we might expect that orbits that are strongly influenced by the BH
might support asymmetric, even lopsided structures. Such appears to be
the case for M31 and NGC4486B, both of which have double nuclei (Lauer
et.al.~1993, 1996); these might be signatures of eccentric nuclear
stellar discs (Tremaine~1995).  A few other galaxies with central
asymmetric light distributions are discussed in Lauer
et.al.~(1995). As Lauer et.al.~(1996) observe, `` A thorough
understanding of the dynamics of eccentric disks might allow us to
estimate the BH mass directly from the disk shape by relating
the scale at which the disc symmetry is broken to the hole mass.'' 

In this paper, we develop a perturbative approach toward
classification of orbits within the BH sphere of influence, taking
advantage of the super--integrability (i.e. degeneracy) of the Kepler
problem. Borrowing an averaging technique from planetary dynamics, we
introduce slow dynamics of precessional motions in \S~2, and discuss
integrability of slow dynamics in two, and three dimensional motions,
for both time independent potentials, as well as the case when the mass
of the BH grows adiabatically.  We introduce a family of planar
potentials that are, in general, cuspy, non--axisymmetric and
lopsided. The harmonic case is particularly suited to simple,
analytic treatment, and we devote \S~3 to an exploration of orbits in
non--axisymmetric, and lopsided cases; orbit families that might have
relevance to double nuclei are briefly discussed. Centred anharmonic
perturbations, with small non--axisymmetry are considered in
\S~4. \S~5 is devoted to outlining the limits of applicability of the
averaging principle, emergence of resonant families, and the outbreak
of chaos. \S~6 provides a summary of our results, paying some
attention to the applicability of the averaging technique to M31 and NGC
4486B.

\section{Slow dynamics}
Consider a cluster of stars, orbiting around a BH located at the 
origin. The motion of a test star is determined by 
the Hamiltonian,
\begin{equation}
\scrh = \frac{v^2}{2} -\frac{GM}{r} + \Phi({\bf r})\,\label{hamil}
\end{equation}
\noindent where ${\bf r}$ and ${\bf v}$ are the position vector and velocity
of the star, $M$ is the mass of the BH, 
and $\Phi$ is the (time independent) mean gravitational
potential of the cluster. 
Within the BH sphere of influence, $\Phi$, by
definition, is a small perturbation. Thus the orbits of stars (that 
are bound to the BH) may be thought of as Keplerian ellipses that
precess and deform on time scales that are slow compared to orbital
times. Hence it is useful to imagine that each star is a slowly
precessing elliptical ring, with its mass distributed inversely
proportional to its speed around its orbit---in the planetary context,
this idea derives from Gauss (see Hagihara~(1971) for a discussion of
Gauss's idea and its applications). Rauch and Tremaine~(1996)
introduced it to explore relaxation effects in star clusters around
supermassive BHs. The elliptical ring approach emerges naturally in
the averaging process  discussed below.
  
When there is a separation of time scales, and one is interested in
the slow evolution, it makes sense to average over high frequency
variations: thus, time averages of physical quantities are equated to
their space averages, an idea that has its roots in the works of
Laplace, Lagrange and Gauss on planetary dynamics. 
Orbit--averaging has also been 
applied to study the evolution of the orbits of comets, perturbed by
the Galactic tidal field (Heisler and Tremaine~1986).
The straightforward  way to achieve 
this is to express
the problem in appropriate action--angle variables, identify the fast
and (the two) slow angles, and integrate over the fast angle. Then the
conjugate (fast) action is predicted to have no secular
evolution. Laplace did this for the solar system, and 
concluded that the semi--major axes of the Keplerian ellipses of the
planets do not evolve secularly. However, this is a reasonable approximation
for the solar system only over times of order $10^4\yr$. A major contribution
to limiting the validity of averaging over orbital motions of the planets are  
mean motion resonances (i.e. ``fast-fast'' resonances). A natural worry is whether 
a similar phenomenon will afflict stellar dynamics in galactic nuclei. Let 
us consider nuclear star clusters in the collisionless limit, when there are 
an infinite number of stars contributing to a total finite mass for the cluster.
Then each stellar orbit can be thought of as a being smoothly populated with stars.
The gravitational force exerted by such an orbit  on a test star will not display
time variation on the orbital time scale (which would have obtained
in the planetary case, when only one planet populates the orbit). Thus we expect 
no mean motion resonances in star clusters around nuclear BHs.\footnote{Of course, 
clusters might have as few as $10^6$ stars, so that each orbit is not smoothly 
populated with stars. Hence there can be some discreteness noise, which would 
make the system somewhat collisional.}  Resonances between fast and slow motions
can be important for orbits with semi-major axes $\sim r_h$, and will stand 
in the way of this naive averaging procedure. Such resonances are the
seeds of chaos, which we explore in \S~5.

To apply the averaging principle, it is convenient to cast the Kepler
problem in terms of the Delaunay action--angle variables: we write
these as $(I, L, L_z; w, g, h)$, where $(I, L, L_z)$ are action
variables, and $(w, g, h)$ are the respective conjugate angle
variables.\footnote{it is unfortunate that the notation employed by
planetary dynamicists overlaps so heavily with those commonly used by
galactic dynamicists for other physical quantities.  Our notation is
non standard, but we hope that it minimizes confusion.}  We list the
basic definitions below, and refer the reader to text books
(c.f. Plummer~1960, Goldstein~1985) for the derivation.
\begin{eqnarray}
I & = & \sqrt{GMa}\,,\quad\mbox{where $a$ is the semi--major axis.}\nonumber \\
w & = & \mbox{mean anomaly; marks time along Keplerian}\nonumber \\
{} & {} & \mbox{orbit---it is $0$ at pericentre, and advances }\nonumber \\
{} & {} & \mbox{by $2\pi$ in one circuit.}\nonumber \\
L & = & \mbox{magnitude of the angular momentum; $L\;\leq\; I\,$.}\nonumber \\
g & = & \mbox{angle from the ascending node to the pericentre.}\nonumber \\  
L_z & = & \mbox{$z$--component of angular momentum.}\nonumber \\
h & = & \mbox{angle from $x$--axis to ascending node.}\label{delaunay}
\end{eqnarray}
\noindent Expressed in the Delaunay variables, the Hamiltonian for the 
Kepler problem (equation~\ref{hamil} with $\Phi$ set equal to zero) assumes
the simple form, $H_{\rm kep}= -(G^2M^2/2I^2)\,$. Thus $w$ advances at a 
uniform rate, 
\begin{equation}
\Omega(I) \;\equiv\;\left(\frac{dw}{dt}\right)_{\rm kep} \;=\;  
\frac{\partial H_{\rm kep}}{\partial I} \;=\; \frac{(GM)^2}{I^3}\,,
\label{omega}
\end{equation}
\noindent whereas the other five Delaunay variables remain constant.

Let us assume that we have managed to express $\Phi$ in terms of the
new variables. Then the full Hamiltonian,
\begin{equation}
\scrh = -\half\left(\frac{GM}{I}\right)^2 + \Phi\,,\label{hamild}
\end{equation}
\noindent where, with some abuse of notation, we now regard
$\Phi$ as a function of all six Delaunay variables.
For a general perturbation, the only conserved quantity is $\scrh$. 
The angles advance at rates given by
\begin{equation}
\frac{dw}{dt}\;=\; \Omega(I) + 
\frac{\partial\Phi}{\partial I}\,,\quad\quad
\frac{dg}{dt}\;=\; \frac{\partial\Phi}{\partial L}\,,\quad\quad
\frac{dh}{dt}\;=\; \frac{\partial\Phi}{\partial L_z}\,.
\label{freqs}
\end{equation}
\noindent When $\Phi$ is small, $\Omega$ is the fastest frequency
in the problem; thus $w$ varies in time much faster than 
$g$ and $h$, and 
averaging simply means that we can integrate the Hamiltonian of 
equation~(\ref{hamild}) over one circuit of $w$---this is 
equivalent to treating a star as an elliptical ring. The averaged Hamiltonian, 
\begin{eqnarray}
\overline{\scrh} & \;=\; & \oint \frac{dw}{2\pi}\,\scrh \;=\;
\oint\,\frac{d\eta}{2\pi}\left(1-\sqrt{1-(L/I)^2}\,\cos\eta\right)\,\scrh
\nonumber \\
& \;=\; & -\half\left(\frac{GM}{I}\right)^2 + \overline{\Phi}\,,
\label{hamilave}
\end{eqnarray}
\noindent governs the slow dynamics of precessional
motions. Since $\overline{\scrh}$ is independent of $w$, we recover
Laplace's result that $I$ is conserved by the slow
dynamics. Furthermore, $\overline{\Phi}$ itself is another slow
integral of motion.\footnote{ $\scrh$ is, of course, exactly
conserved, but this does not give us any extra conserved quantities
after averaging.} Without further ado, we can reach some general
conclusions about slow dynamics in the region $r\ll r_h\,$. Below we
list our conclusions in order of increasing generality:

\begin{itemize}
\item For razor--thin discs, motion is confined to the $z=0$ plane.
Any $\Phi(x, y)$ has the two slow integrals of motion, $I$ and
$\overline{\Phi}$. Hence the slow dynamics for two dimensional
potentials, however non--axisymmetric or lop--sided, is integrable,
and a straightforward classification of orbits is possible.
 
\item For axisymmetric $\Phi$ in three dimensions, $L_z$ is 
an exact integral of motion (since $\Phi$ is independent of $h$). We
now have three integrals of motion ($L_z$, $I$ and $\overline{\Phi}$),
and the slow dynamics is again fully integrable.

\item In three dimensions, when $\Phi$  has
no symmetry whatsoever, we still have the integrals $I$ and
$\overline{\Phi}$. The slow motion can be chaotic, but it is clear
that the chaos is limited.
\end{itemize}

\subsection{Adiabatic growth of the BH}

Averaging is also applicable to time dependent perturbations,
$\Phi({\bf r}, t)$, when the time variations are slower than the orbital
times; as before, $I=\sqrt{GMa}$ is a quasi--invariant.  In some
scenarios of the formation and subsequent growth of the BH, its mass
increases adiabatically (Peebles~1972, Young~1980); conservation of
$I$ implies that the semi--major axis shrinks in proportion to the growth
of $M$.  When the growth time is also much longer than the
precessional timescales, additional (adiabatic) invariants might
arise. These will, in general, be related to the actions corresponding
to precessional degrees of freedom $(L, L_z, g, h)$. We again consider
cases with different spatial symmetry.

\begin{itemize}
\item For razor--thin discs, the conserved actions are $I$ and $\oint L\, dg$, 
where the integral is performed at fixed time, over one cycle of motion
in the $L-g$ plane.

\item For three dimensional, axisymmetric cases, in addition to 
$I$ and $\oint L\, dg$, we have $L_z$ as an exactly conserved quantity.

\item For configurations that have no spatial symmetry, $I$ is in
general the only conserved quantity, since resonances or precessional
chaos might destroy the other adiabatic invariants.
\end{itemize}
\noindent
It is difficult to make further progress without taking into account
the self--consistent evolution of the cluster potential.

\subsection{Slow planar dynamics}

For razor--thin discs, orbits are restricted to the $x-y$ plane. $(I,
w)$ determine the semimajor axis and position on the orbit, and $(L,
g)$ the eccentricity and orientation of the ellipse in the plane. The
only difference from the three dimensional case is that we allow $L$
to take both $\pm$ values.  It proves convenient to set $w=\eta
-e\sin\eta$, where $\eta$ is the eccentric anomaly, and $e$ is the
eccentricity given by $e^2=[1-(L/I)^2] \leq 1$. If we imagine
cartesian coordinates, $(x^{'}, y^{'})$, centred at the origin, such
that positive $x^{'}$ is along the major--axis toward the pericentre,
and $y^{'}$ is parallel to the minor--axis, we have $x^{'}=a(\cos\eta
- e)$, and $y^{'}=a\sqrt{1- e^2}\sin\eta$. The primed coordinates
being rotated by angle $g$ with respect to the fixed coordinates, $(x,
y)$, we obtain
\begin{eqnarray}
x &=& a\left\{{\cos g}\,\left(\cos\eta - \sqrt{1-\ell^2}\right) -\ell{\sin g}\,\sin\eta\right\}
\nonumber \\[1em]
y &=& a\left\{{\sin g}\,\left(\cos\eta - \sqrt{1-\ell^2}\right) +\ell{\cos g}\,\sin\eta\right\}\,,
\label{xy2del}
\end{eqnarray}
\noindent where $\ell=L/I$ is a dimensionless angular momentum which
takes values, $-1\leq\ell\leq 1\,$.

The frequency of apsidal precession of a test particle 
will depend on both the size ($a$) and shape ($e$) of the 
orbit. We wish to associate a characteristic value of this 
frequency, $\mu$, with the size (but not the shape) of the orbit. 
Hence it proves convenient to take $\mu$ to be of order the precession 
frequency for circular orbits (a somewhat paradoxical notion, but the limit is
sensible). Then the ratio, $\delta(I)\equiv\mu(I)/\Omega$ is
a convenient small parameter for slow dynamics. We now use a
dimensionless time, $\tau=\mu t$, as an appropriate temporal measure for 
slow dynamics. Defining the dimensionless, averaged Hamiltonian,
\begin{equation}
H(\ell,g; I)=\frac{\overline{\Phi}}{I\mu}
=\oint\,\frac{d\eta}{2\pi}\left(1-\sqrt{1-\ell^2}\cos\eta\right)
\left(\frac{\Phi}{I\mu}\right)
 \,,\label{slowham}
\end{equation}
\noindent the equations of motion take the standard form,
\begin{equation}
\dot{\ell}\equiv\frac{d\ell}{d\tau}=-\frac{\partial
H}{\partial g}\,,\quad\quad
\dot{g}\equiv\frac{dg}{d\tau}=\frac{\partial
H}{\partial\ell}\,.\label{sloweqns}
\end{equation}
\noindent In fact, much of the qualitative picture of the 
orbit families can be gleaned by studying the contour plot of
$H$ in the $(g,\ell) $ plane (for some value of the constant $I$). 
 
\subsection{A family of planar model potentials}
To develop some analytic understanding of orbital structure, we
introduce a family of planar potential perturbations that allow for
both non--axisymmetry and lopsidedness.\footnote{In the lopsided
cases, the potential exerts a force on the BH, so we should imagine
that the BH is `nailed' down. It is possible to improve the situation
by including cubic and higher order terms in the potential, but at the
cost of considerably more complexity. We decided to forgo this luxury,
since our primary goal in this paper is to pick out interesting
orbits.}  Let us define a quadratic form,
\begin{equation}
q^2\;=\; (x-d_1)^2 + \frac{(y-d_2)^2}{b^2}\,,\label{eqn.q2}
\end{equation}
and the family of planar potentials,
\begin{equation} 
\Phi(x,y)=\left\{ \begin{array}{ll}
\mbox{sgn}\,(\alpha)\Phi_0\,r_0^{-\alpha}
\left(q^2 + r_c^2\right)^{\alpha/2}\,, & \alpha\neq 0\,; \\[1em]
(\Phi_0/2)\log\left(q^2 + r_c^2\right)\,, & \alpha=0\,,
\end{array} \right.
\label{potential.2}
\end{equation}
\noindent The potentials depend on the five structural parameters,
$(\,{\bf d}, r_c, b^2, c^2, \alpha \,)$, and a magnitude parameter
$(\Phi_0\,r_0^{-\alpha})$.  The potentials are centred at a point that
is displaced from the BH by ${\bf d}\equiv(d_1,d_2)$.  Potential
isocontours have a core of radius $r_c$, and constant axis ratio,
$b$. The exponent $\alpha$ is a measure of the potential gradient
across the contours; we require $2\geq\alpha > -1$, a range that
includes homogeneous cores as well as very centrally concentrated mass
distributions.  The potentials are lopsided for non zero ${\bf d}$,
and are cuspy when $r_c=0$. When both ${\bf d}={\bf 0}$ and $r_c=0$,
the dynamics itself is scale--free, so we may set $a=1\,$.  Lopsided
potentials are not scale--free, even when $r_c=0$; this is because the
lopsidedness sets a scale $(d)$. The magnitude,
$(\Phi_0\,r_0^{-\alpha})$ measures the slowness of the
dynamics. Specifically, we choose units such that
$\mu=(\Phi_0a^\alpha/Ir_0^\alpha)= (\Phi_0a^\alpha/\sqrt{GMa}\,
r_0^\alpha)$. As discussed above, the small parameter is $(\mu/\Omega)
\propto a^{(1+\alpha)}$; since $\alpha > -1$, averaging is a very good
approximation for small $a$.

\section{Planar harmonic perturbations}

The slow dynamics of {\em harmonic} perturbations, for which
$\alpha=2$, can be studied exactly for arbitrary non--axisymmetry, and
lopsidedness.  Pfenniger and de~Zeeuw~(1989) studied stellar orbits in the  
combined fields of a point mass, and a non-axisymmetric harmonic potential. 
For small energies they report that the orbits are essentially regular orbits 
in the Kepler potential of the point mass. As we demonstrate in this section, 
this is not quite true; the lenses and loops that emerge from our investigations
are only {\em locally} Keplerian. They are better described as precessing 
and deforming Keplerian ellipses. They point out (correctly)
the chaos that occurs for orbits with $a\sim r_h$, and discuss  
orbits of much higher energies. These are the orbits that will determine the 
overall shape of a galaxy, and they make the interesting point that, if 
the $x$ and $y$ frequencies of the harmonic potential are commensurate, then 
the high energy orbits will tend to stay well away from  the centre, and hence
avoid being strongly perturbed by the BH. We do not discuss their work any further, 
because our interest is dynamics within $r_h$.

The core radius loses dynamical significance in this
case, so in our explorations of harmonic perturbations, we set
$r_c=0\,$:
\begin{equation}
\frac{\Phi}{I\mu}= \frac{1}{a^2}\left(d_1^2 +\frac{d_2^2}{b^2}
-2d_1x -2\frac{d_2y}{b^2} + x^2 + \frac{y^2}{b^2}\right)\,.
\label{harmonic.2}
\end{equation}
\noindent Substituting for $x$ and $y$, the expressions given in
equations~(\ref{xy2del}), gives us the perturbation explicitly
in terms of the Delaunay variables. We then average over $w$, and 
obtain the following expression for the slow Hamiltonian:\footnote{
A term, $\left[(5/2) +
(5\epsilon/4) + (d_1/a)^2 + (1+\epsilon)(d_2/a)^2\,\right]$,
has been dropped in the slow Hamiltonian, because it is does not 
contribute to slow dynamics.}
\begin{eqnarray}
H & = & -\frac{3}{2}\left(1+\frac{\epsilon}{2}\right)\ell^2 \;-\;
\frac{5\epsilon}{4}\left(1-\ell^2\,\right)\cos 2 g\nonumber \\
&{}& +\; 3\sqrt{1-\ell^2}\left(\frac{d_1}{a}\cos g + (1+\epsilon)
\frac{d_2}{a}\sin g\,\right)\,, 
\label{ave.harmonic}
\end{eqnarray}
\noindent where we have introduced $\epsilon=(b^{-2} -1 )$ as a measure
of non--axisymmetry. The parameter space is 3 dimensional, $(\epsilon, 
\;d_1/a\;, \;d_2/a\;)$, so that the slow dynamics of even planar, harmonic 
perturbations is quite rich. Below we study a few cases. 

\subsection{Centred harmonic perturbation: $d_1=d_2=0$}

The simplest case is a centred ($d_1=d_2=0\,$) perturbation, for which
we may without loss of generality, set $\epsilon\geq 0\,$.  The
Hamiltonian is
\begin{equation}
H=-\frac{3}{2}\left(1+\frac{\epsilon}{2}\right)\ell^2 
\;-\; \frac{5\epsilon}{4}\left(1-\ell^2\,\right)\cos 2 g\,, 
\label{cent.harmonic}
\end{equation}
\noindent and the equations of motion are,
\begin{eqnarray}
\dot{\ell} &=& -\frac{\partial H}{\partial g} \;=\; -\frac{5}{2}
\epsilon\,\left(1-\ell^2\right)\sin\,2 g \nonumber \\
\dot{ g} &=& \frac{\partial H}{\partial\ell} \;=\;
-3\left(1+\frac{\epsilon}{2}\right)\ell \;+\; \frac{5\epsilon}{2} 
\ell \cos\,2 g\,,\label{hcent.eom}
\end{eqnarray}

\begin{figure}
\centerline{\hbox{\epsfxsize=1.7in\epsfbox{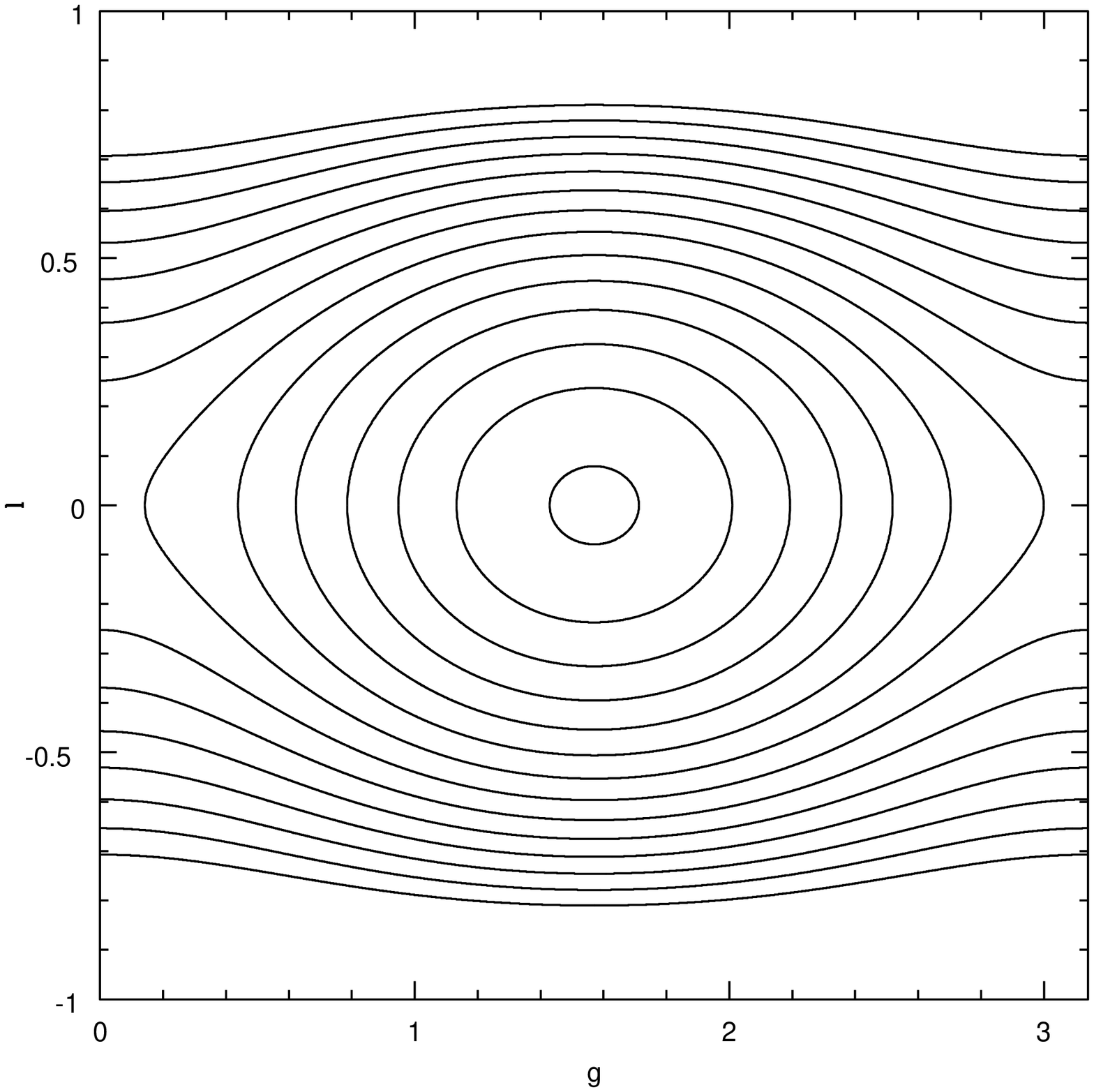}
                  \epsfxsize=1.7in\epsfbox{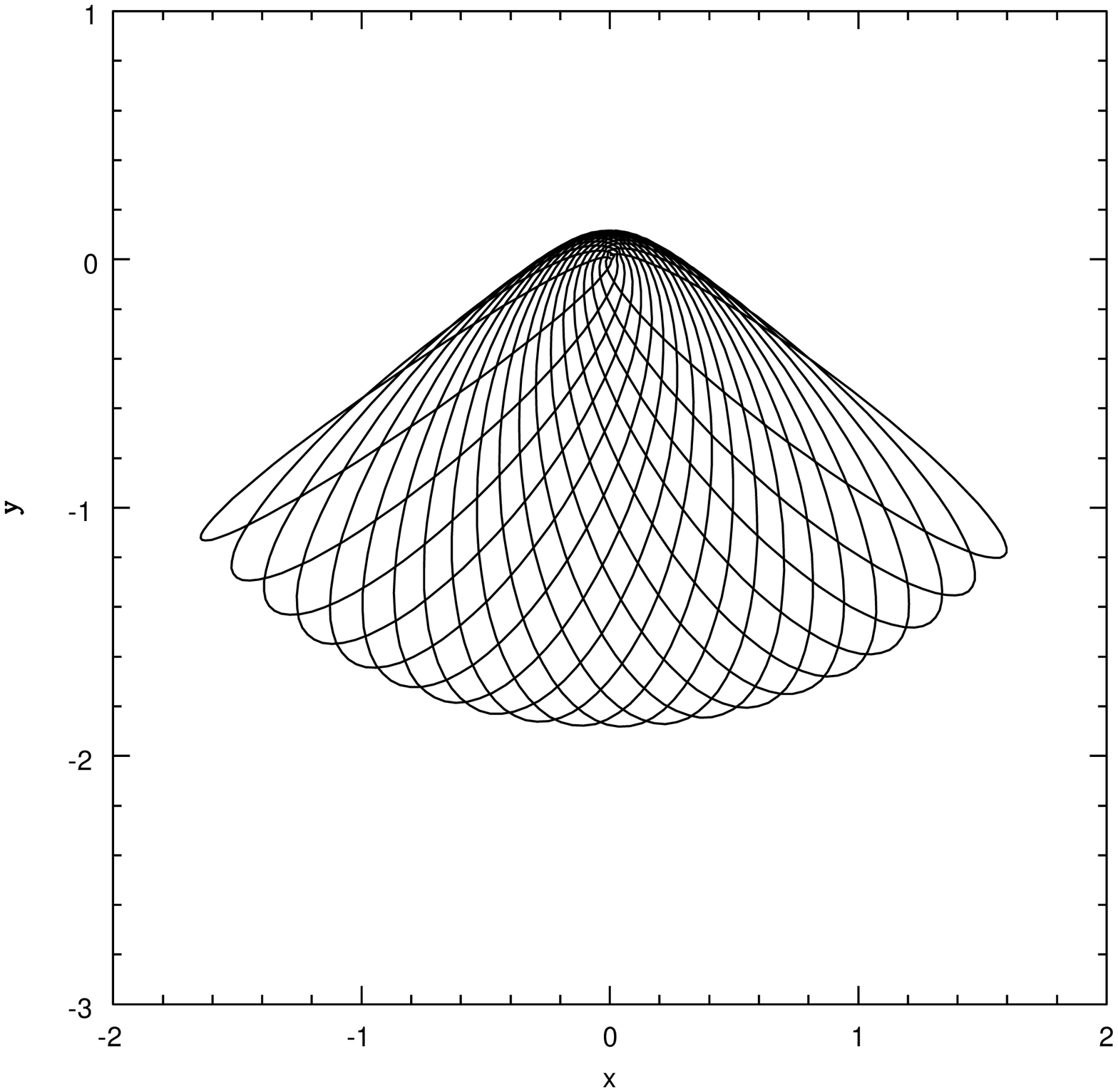}}}
\centerline{\hspace*{0.6in}$(a)$\hfill$(b)$\hspace{0.5in}}
\centerline{\hbox{\epsfxsize=1.7in\epsfbox{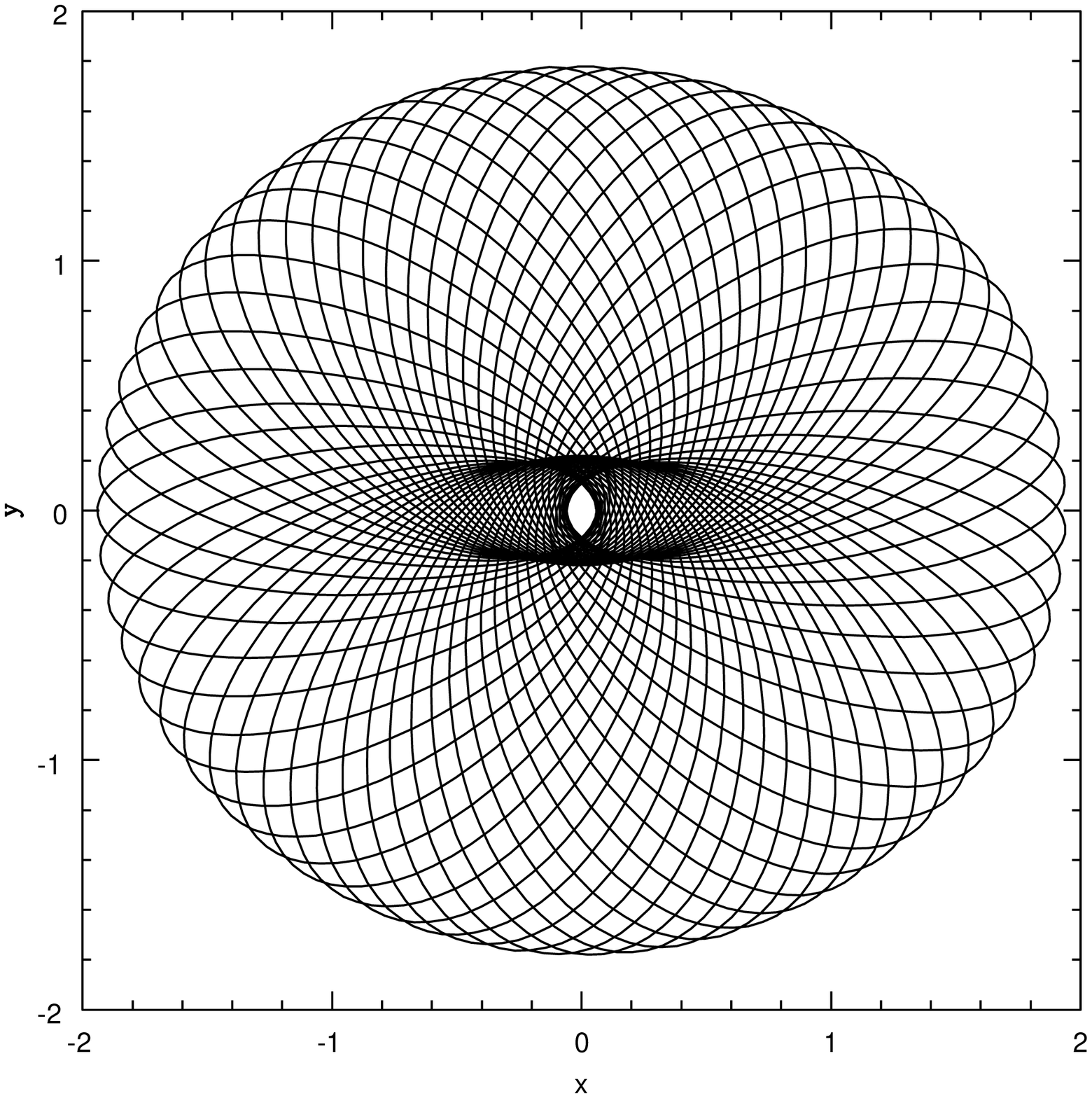}
                  \epsfxsize=1.7in\epsfbox{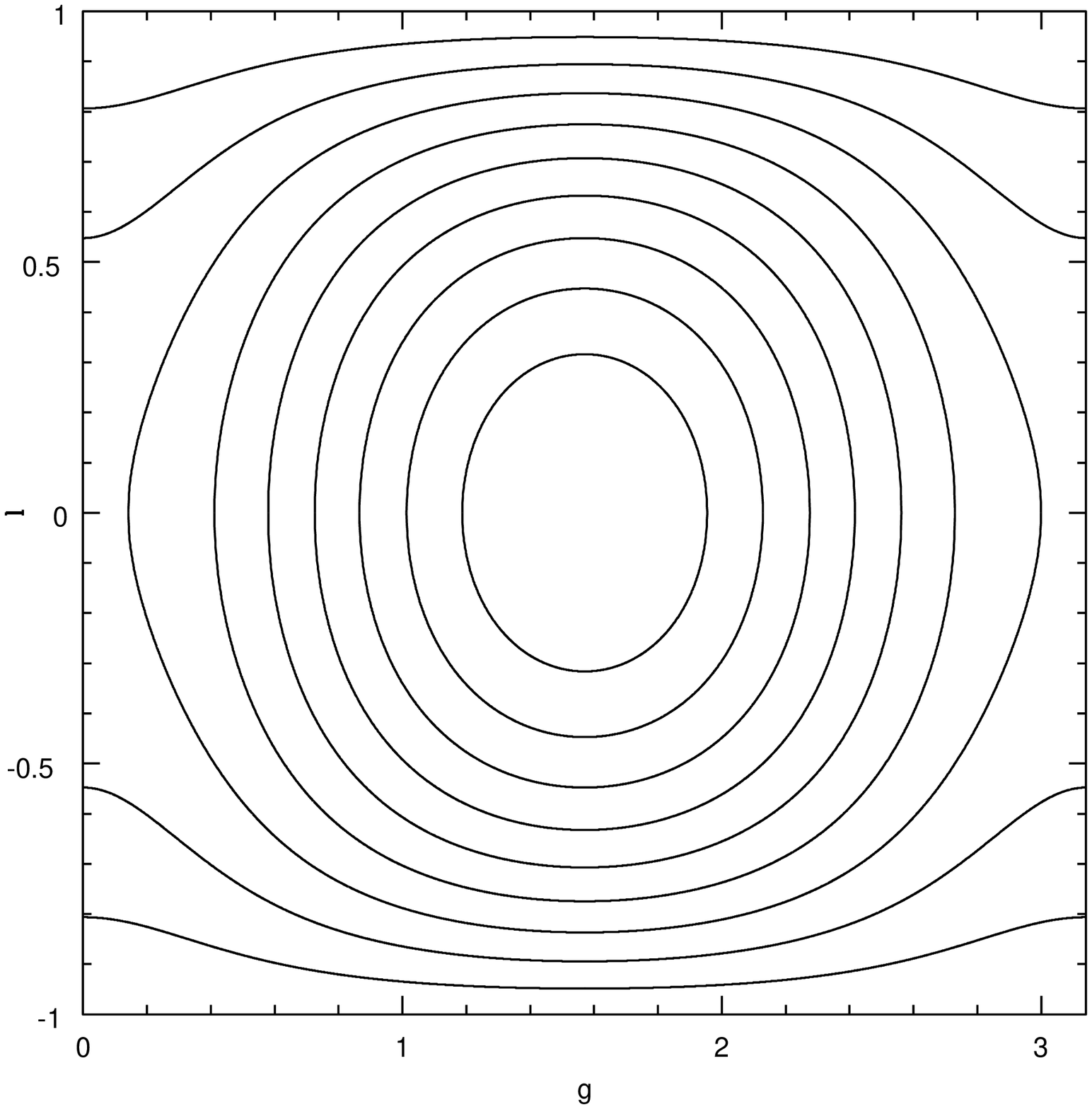}}}
\centerline{\hspace*{0.6in}$(c)$\hfill$(d)$\hspace{0.5in}}
\caption[Figure 1]{Planar, centred, harmonic perturbation: (a)
isocontours of $H$ in the $( g,\ell)$ plane for $\epsilon=0.25$, (b) a
lens orbit in real space corresponding to $H=-0.13$, (c) a loop orbit
in real space corresponding to $H=-0.47$, (d) isocontours of $H$ in
the $( g,\ell)$ plane for $\epsilon=1$}
\label{fi:level}
\end{figure}

\noindent When $\epsilon=0$, the full 
Hamiltonian is axisymmetric and the angular momentum is an exactly conserved
quantity. The orbits are the rosette--like figures given in standard 
text books on classical mechanics. The averaged description, of course, 
coincides with this picture; $H=-3\ell^2/2\,$,  the isocontours of 
which are simply lines of constant $\ell$ in the $( g,\ell)$ plane. 
These {\em loop} orbits are now viewed as Keplerian ellipses that precess 
at a rate, $\dot{ g}=-3\ell\,$.

When $\epsilon\neq 0$, the non--axisymmetry gives birth to a new
family of orbits, the {\em lenses}. Figures~1a shows contour plots of
$H$ for the case $\epsilon=0.25$, for which the isocontours of the
perturbation have axis ratio $b\simeq 0.9\,$. The lenses are parented
by the short axis orbits, which appear as the stable fixed points
located at $(\pi/2, 0)$ and $(3\pi/2, 0)$.  Figure~1b shows a lens
orbit (for $H=-0.13$); $\ell$ oscillates about zero, whereas $ g$
librates about the short axis.  When $g=\pi/2$, it is maximally round, 
and when the pericentre reaches
its maximum deviation from the short axis, the lens elongates to a
line (``radial orbit''); its angular momentum now switches sign, and
the orbit returns to maximal roundness at $ g=\pi/2$, but with the
opposite sign of $\ell\,$.  Unlike lenses, loops have a definite sign
for $\ell$, and $ g$ circulates through $2\pi$ in one period. The
loop in Figure~1c is a deformed rosette that spends somewhat more time
near the long axis, as it circulates around the origin. The unstable
fixed points at $(0, 0)$ and $(\pi, 0)$ correspond to the long axis
orbits. Loops are separated from lenses by the separatrix that
straddles the unstable fixed points. The separatrix orbit, as well as
the loops and lenses that are close to it spend a lot of time in the
vicinity of the long axis. As $\epsilon$ increases, lenses occupy an
increasingly larger fraction of phase space (see
Figure~1d). 

\begin{figure}
\centerline{\hbox{\epsfxsize=1.7in\epsfbox{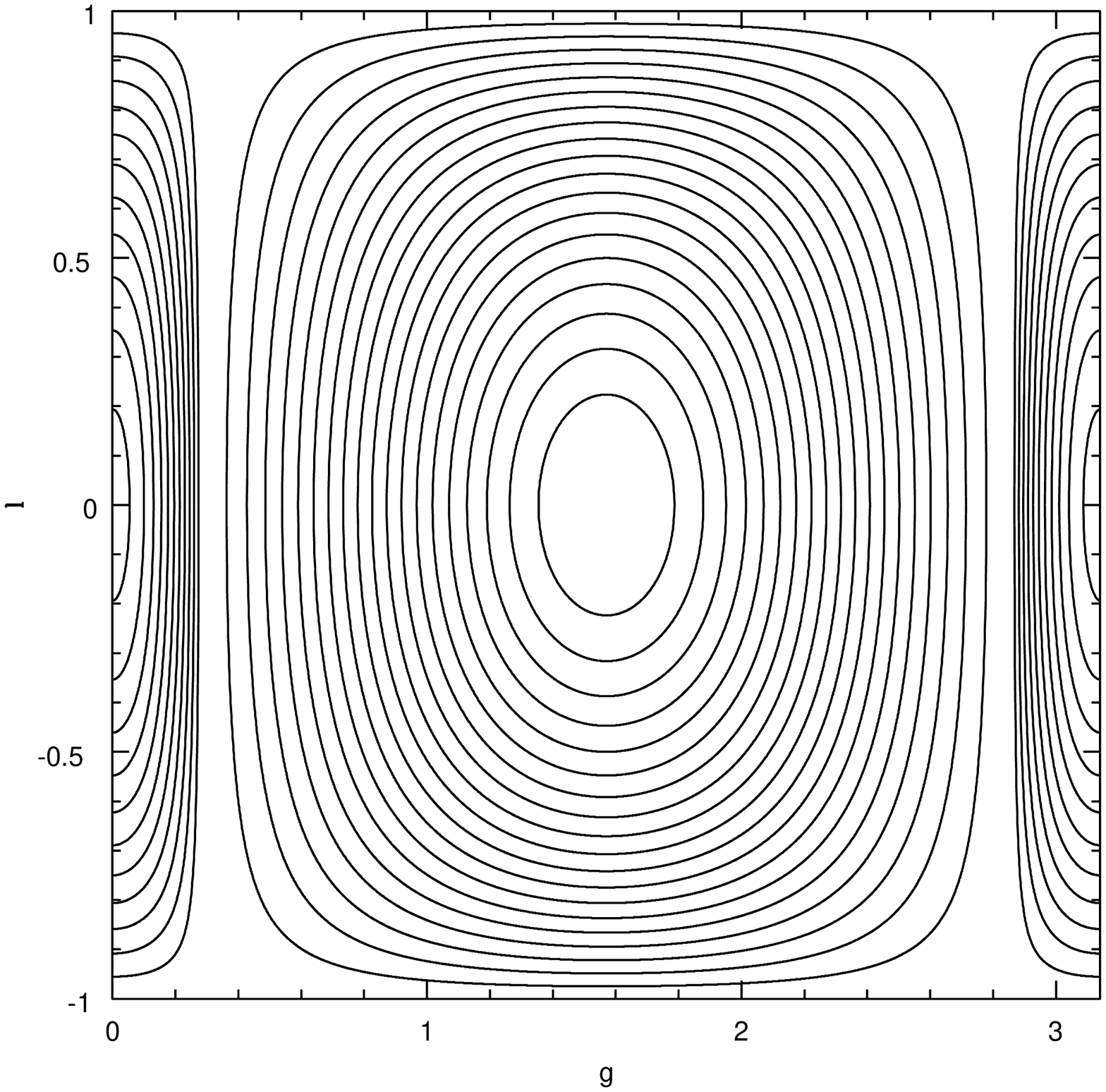}
                  \epsfxsize=1.7in\epsfbox{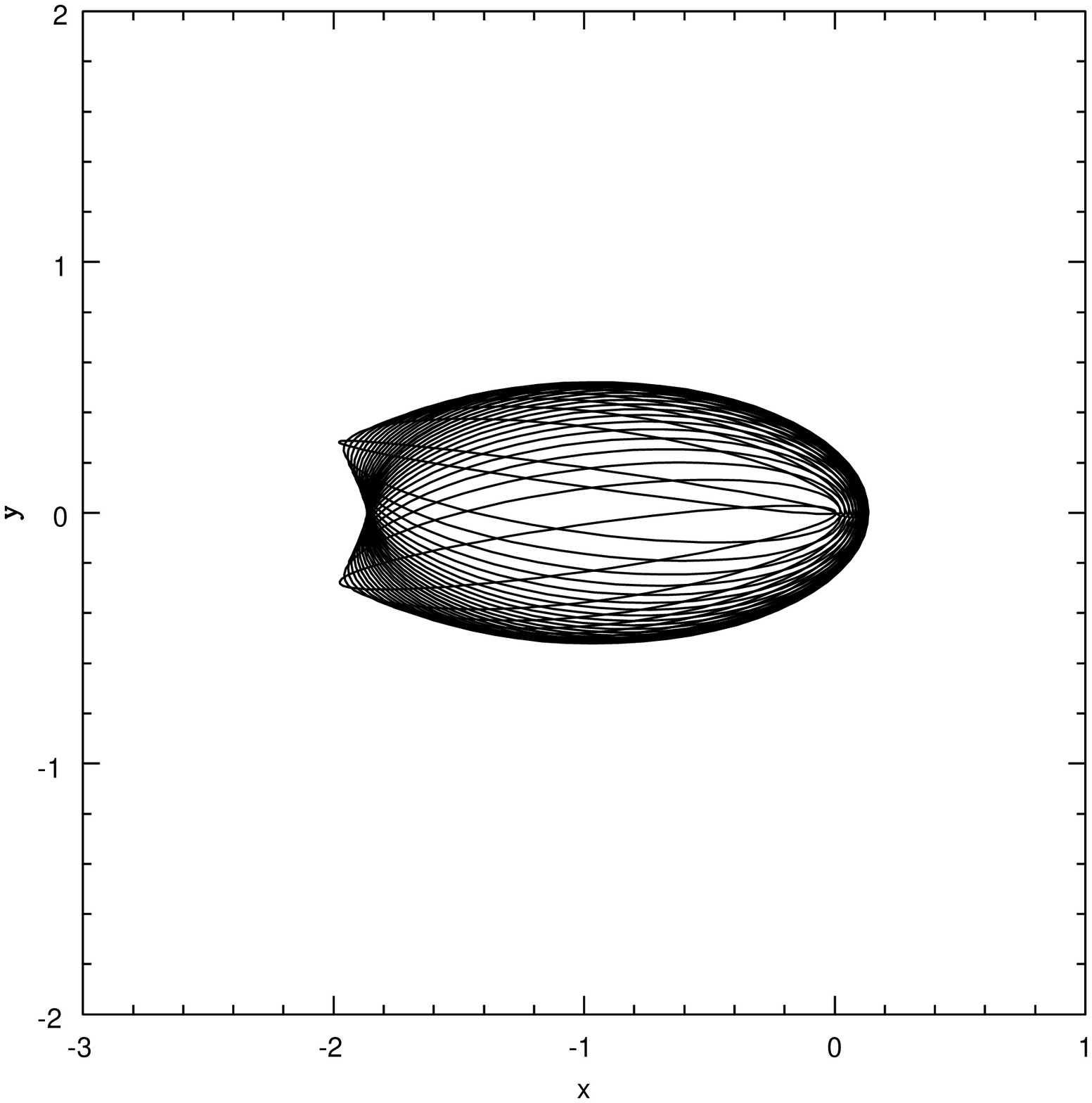}}}
\centerline{\hspace*{0.6in}$(a)$\hfill$(b)$\hspace{0.5in}}
\caption[Figure 2]{Planar, centred, harmonic perturbation: (a) isocontours of $H$ 
for $\epsilon=5.0$, (b) A long--axis lens orbit, for $H=-6.0$.} 
\label{fig:longaxis}
\end{figure}

When $g=n\pi$, equations~(\ref{hcent.eom}) imply that $\dot{\ell}=0$
and $\dot{ g}= (\epsilon - 3)\ell$.  Thus, when $\epsilon=3$, $ g=0,
\pi$  are {\em lines} of fixed points in the $( g,\ell)$ plane. In fact, 
when $\epsilon=3$, the region occupied by the short--axis lenses has
expanded to its fullest, squeezing the loops out of existence. This
value of $\epsilon$ corresponds to axis ratio of two (for the
isocontours of the harmonic perturbation), a case for which the full
(unaveraged) Hamiltonian is exactly integrable (see Appendix~A1 for details). 
For $\epsilon >3$, loops are completely absent, and all of phase space is
populated by lenses (see Figure~2a). The long--axis orbits now become
stable, and parent families of {\em long--axis lenses}, one of which
is shown in Figure~2b. We can demonstrate this change in stability of
the long--axis orbit, by expanding the $H$ of
equation~(\ref{cent.harmonic}) to lowest order about the fixed point
$(0,0)$:
\begin{equation}
H \;=\; -\frac{5}{4}\epsilon + (\epsilon -3)\,\frac{\ell^2}{2} + 
\frac{5\epsilon}{2}\, g^2\,.\label{stab.lao}
\end{equation}
\noindent When $\epsilon<3$, the coefficients of $\ell^2$ and $ g^2$
are of opposite signs (the long--axis orbit is unstable), and have the
same sign for $\epsilon>3$ (when the long--axis orbit is stable).  We
can also see that the short--axis orbit remains stable by expanding
$H$ about $(\pi/2,0)$.  With $ g=\pi/2 + \delta g$,
\begin{equation}
H \;=\; \frac{5}{4}\epsilon -(4\epsilon +3)\,\frac{\ell^2}{2}- 
\frac{5}{2}\epsilon\,(\delta g)^2\,.\label{stab.sao}
\end{equation}
\noindent The coefficients of both $\ell^2$ and $(\delta g)^2$ have
the same sign, so the short--axis orbit is always stable.

\subsection{Lopsided harmonic perturbation: $d_1\neq 0$, $d_2=0$, $\epsilon=0$}
New orbits emerge when the centre of the perturbation is displaced
from the BH.  The simplest case occurs for $\epsilon=0$, for which the
isocontours of the perturbing potential are circles. In this case we
may, without loss of generality, set $d_2=0$, and let $d_1=d\,$. The
Hamiltonian,
\begin{equation}
H\;=\; -\frac{3}{2}\,\ell^2 \;+\; 3\,\frac{d}{a}\sqrt{1-\ell^2}\cos g
\,.\label{lopsided}
\end{equation}
\noindent now has a $(\cos g)$ term, which distinguishes between
orbits that are {\em aligned} with, from those that are {\em
anti--aligned} with the lopsidedness of the perturbation. We recall
that $ g$ is the angle from the major ($x$) axis to the pericentre of
the instantaneous Keplerian ellipse. Thus orbits with $ g=\pi$ are
aligned, whereas $ g=0$ are anti--aligned with the lopsidedness. The
only free parameter in $H$ is $d/a$, which may be thought of as a
dimensionless measure of lopsidedness. Below we explore the orbital
structure as a function of this parameter.

The panels in Figures~3 show the dependence of the isocontours of $H$
on lopsidedness. Changes in the topology of isocontours occur when
$d/a$ crosses $0.5$ (Figure 3a to 3b), and again when $d/a$ crosses
unity (Figure 3c to 3d). This behavior is best understood by following
the location and stability of the fixed points. The equations of
motion are,
\begin{eqnarray}
\dot{\ell} &=& -\frac{\partial H}{\partial g} \;=\; 3\frac{d}{a}
\sqrt{1-\ell^2}\,\sin g \nonumber \\
\dot{ g} &=& \frac{\partial H}{\partial\ell} \;=\;
-3\ell \;-\; 3\,\frac{d}{a}\,\frac{\ell}{\sqrt{1-\ell^2}}\,\cos g\,,\label{eom}
\end{eqnarray}
\noindent and the fixed points are determined by requiring $\dot{\ell}=
\dot{ g}=0\,$. There are four fixed points, located at 
\begin{eqnarray}
 g &\;=\;&   0\,,\quad \ell \;=\; 0\,,\quad\quad\mbox{(anti--aligned radial
orbit)};\nonumber \\[1ex]
 g &\;=\;& \pi\,,\quad \ell \;=\; 0\,,\quad\quad\mbox{(aligned radial
orbit)};\nonumber \\[1ex]
 g &\;=\;& \pi\,,\quad \ell \;=\; \pm\,\sqrt{1-\frac{d^2}{a^2}}\,,
\quad\quad\mbox{(aligned loops)}.
\label{fxdpts}
\end{eqnarray}
\noindent 

As  Figures~3 show, the anti--aligned, radial orbit
is a stable fixed point. The stability can be established by
expanding $H$ to lowest order about $(0,0)$:
\begin{equation}
H \;=\; 3\frac{d}{a} -\frac{3}{2}\left(1+\frac{d}{a}\right)\,\ell^2
 -\frac{3}{2}\frac{d}{a}\, g^2 +\ldots\,.\label{h.aaro}
\end{equation} 
\noindent The coefficients of $\ell^2$ and $ g^2$ have the same sign,
so the fixed point is stable. 

\begin{figure}
\centerline{\hbox{\epsfxsize=1.7in\epsfbox{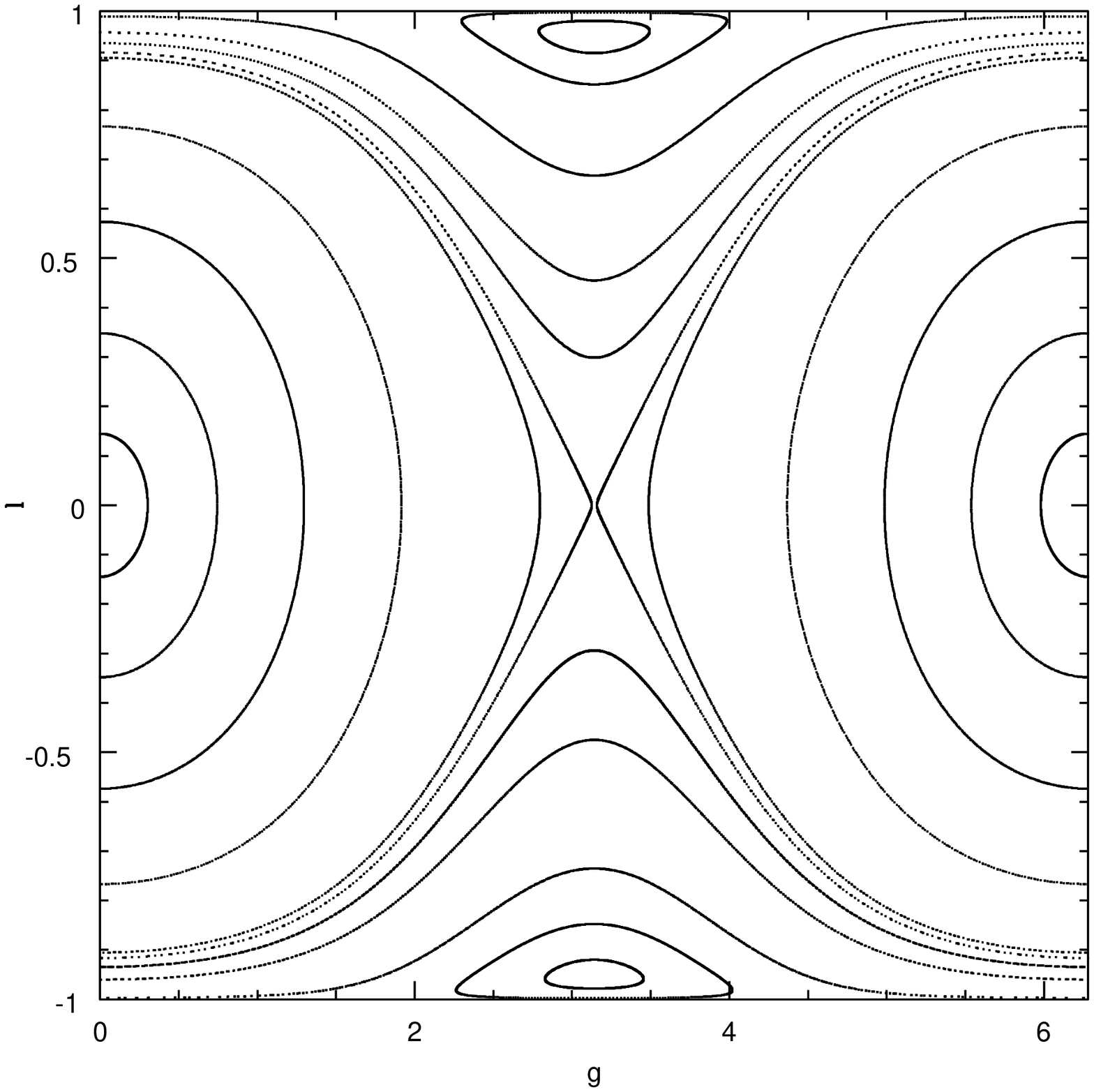}
                  \epsfxsize=1.7in\epsfbox{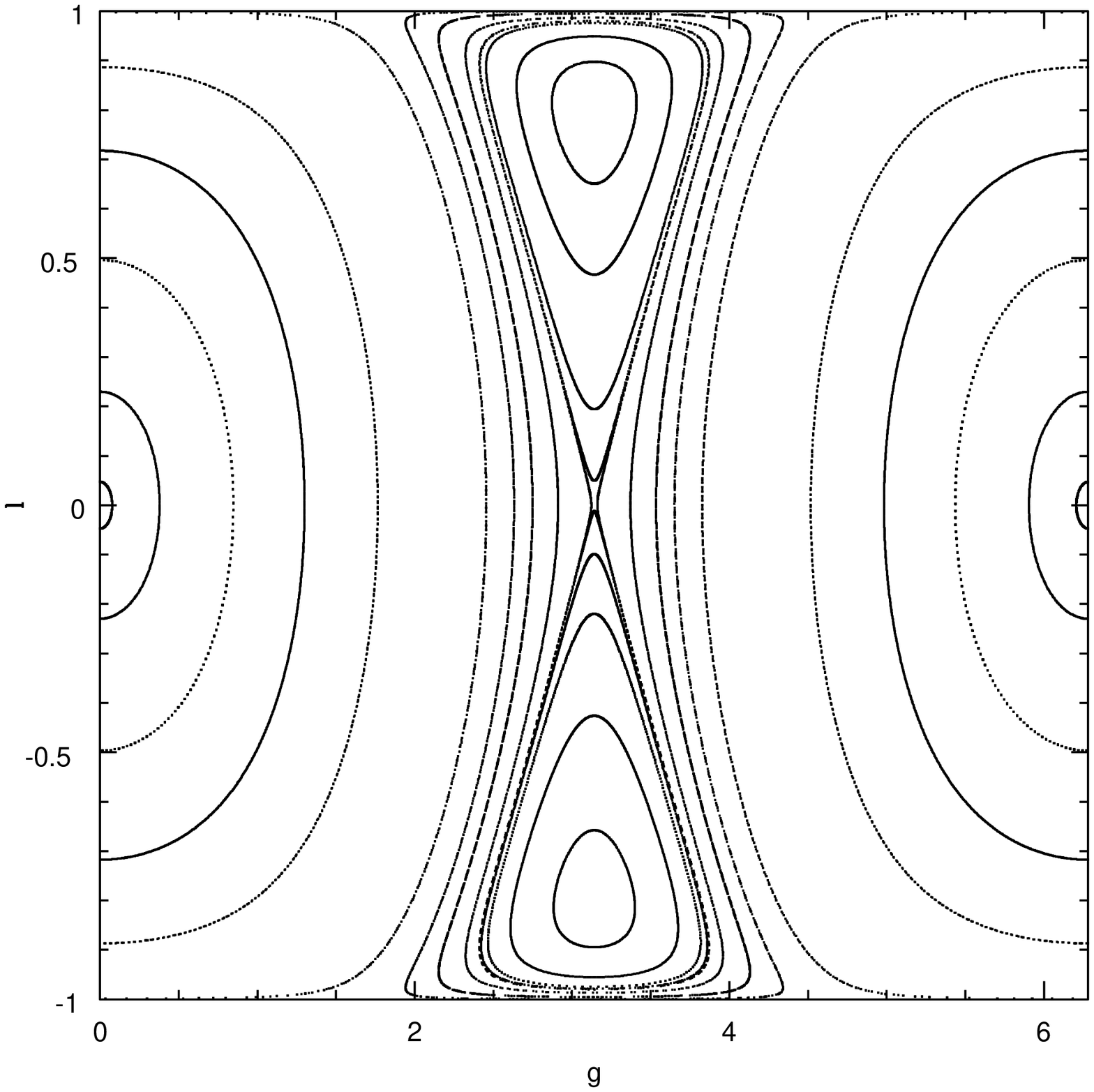}}}
\centerline{\hspace*{0.6in}$(a)$\hfill$(b)$\hspace{0.5in}}
\centerline{\hbox{\epsfxsize=1.7in\epsfbox{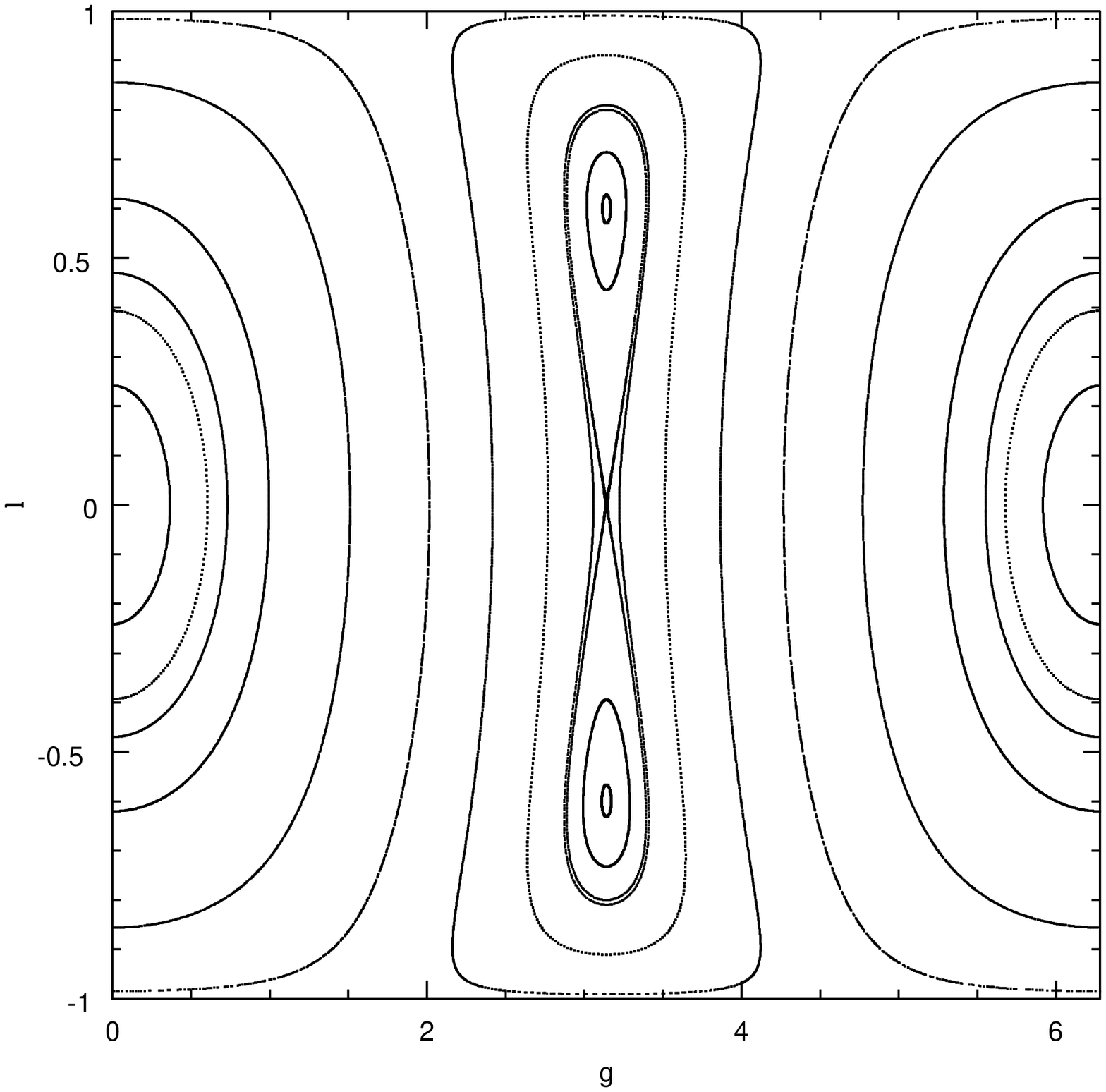}
                  \epsfxsize=1.7in\epsfbox{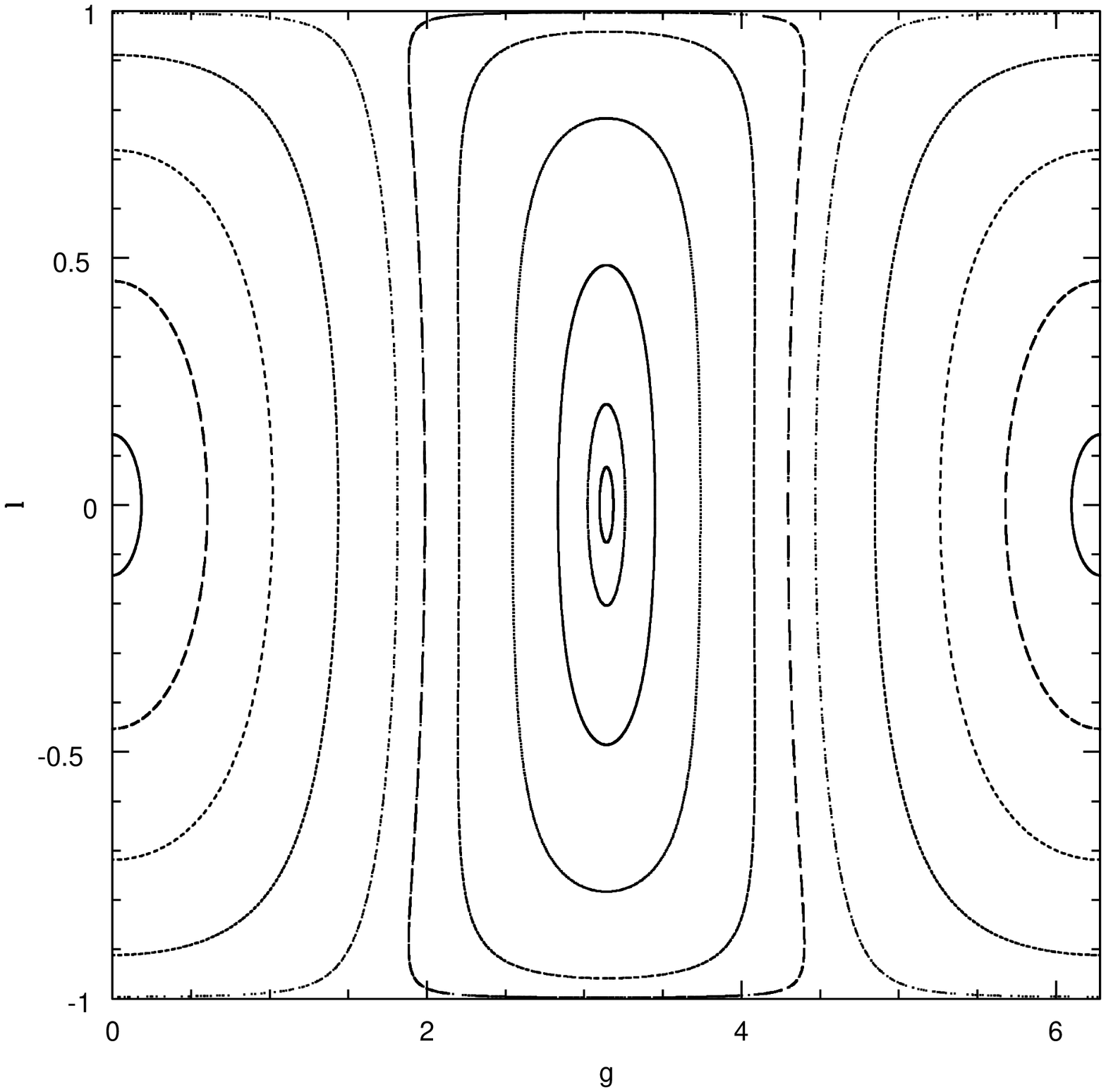}}}
\centerline{\hspace*{0.6in}$(c)$\hfill$(d)$\hspace{0.5in}}
\caption[Figure 3]{Isocontours of $H$ for  planar, lopsided, 
harmonic perturbation (a) $d=0.3 a$, (b) $d=0.6 a$, (c) $d=0.8 a$,
(d) $d=1.5 a$.}
\end{figure}

When $d < 0.5 a$ (see Figure 3a), the region of anti-aligned
lenses, shares a border (a ``separatrix'' which goes through 
the origin of the $\ell-g$ plane) with rosette--like loops. 
These loops have a definite sense of circulation, anti--clockwise or 
clockwise accordingly as $\ell$ is positive or negative. 
As the average value of $|\ell|$
increases, we sample this family of loops until we reach that
rosette which at $g=0$ has $|\ell|= 1$. 
This is a critical orbit, beyond
which we enter the region occupied by the more interesting family of
aligned loops, a family which we explore further down. As $d$
increases to $0.5 a$, the separatrix, which keeps apart anti-aligned lenses
from rosettes, grows to its maximum allowable half-width of $1$,
squeezing the rosettes out of existence. As can be seen in Figure 3b,
(for which $d/a=0.6$), the rosettes have disappeared. The
other fixed points correspond to orbits that are aligned with
the lopsidedness. The aligned loops have eccentricity, $e=d/a\,$; they
begin as circular orbits when $d=0$, elongate with increasing
lopsidedness, reducing to radial orbits when $d=a$. As discussed above, when
$d/a < 0.5$, 
the aligned loops are separated from the anti-aligned lenses by the
rosettes.  As $d/a$ increases beyond $0.5$, and while $d/a < 1$, the
family of aligned loops shrinks, and is now separated from the
anti-aligned lenses by lens-like orbits surrounding the aligned radial
orbit.  All through this variation of $d$ from $0$ to $d=a$, the
aligned radial orbit is unstable. At $d=a$, the two aligned loops
(with either sign of $\ell$) merge with the aligned radial orbit,
which now remains stable for $d>a\,$. Again, this change of stability
of the aligned, radial orbit may be verified by expanding the
Hamiltonian about $(\pi, 0)$. Setting $ g= \pi + \delta g\,$, to
lowest order,
\begin{equation}
H \;=\; -3\frac{d}{a} \;-\; \left(1-\frac{d}{a}\right)\,\frac{3\ell^2}{2}
\;+\; \frac{3}{2}\frac{d}{a}\,(\delta g)^2 +\ldots\,.\label{h.aro}
\end{equation}

\noindent When $d<a$, the coefficients of $\ell^2$ and $(\delta g)^2$
have opposite signs (fixed point unstable, but have the same sign for
$d>a$ (fixed point stable). Hence there are three kinds of fixed points:
 
\noindent 1.\quad The anti--aligned radial orbit is  stable for all $d/a$. This
parents a family of {\em anti--aligned lenses} (see Figure~4a).

\noindent 2.\quad The aligned radial orbits are stable when $d>a$,
and parents a family of {\em aligned lenses} (see Figure~4d).

\noindent 3.\quad The aligned loop orbits are stable when $d<a$. These
parent a family of {\em librating loops}, to be discussed in some detail
below. 

There are no orbits that correspond to anti--aligned loops. 

\begin{figure}
\centerline{\hbox{\epsfxsize=1.7in\epsfbox{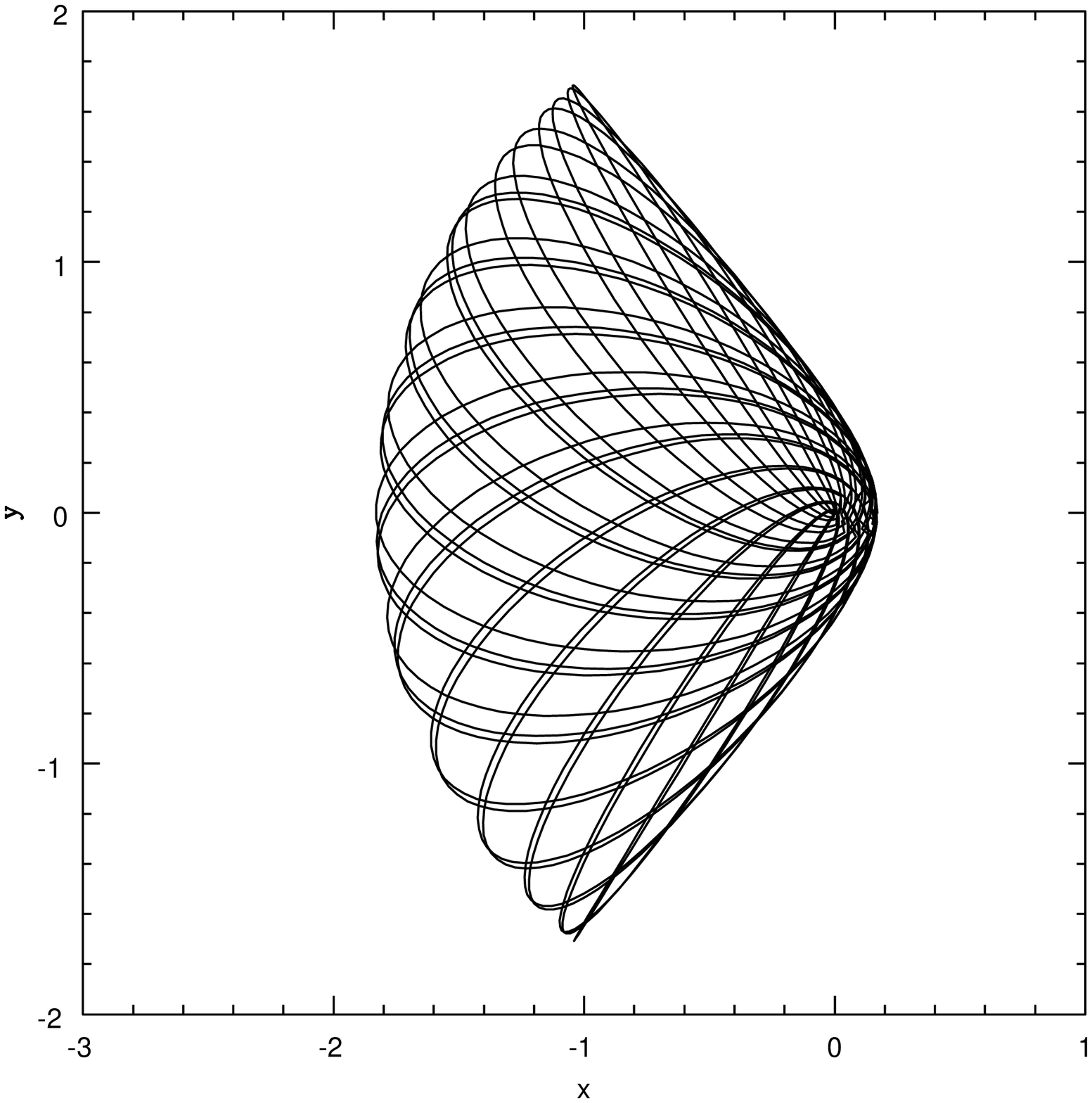}
                  \epsfxsize=1.7in\epsfbox{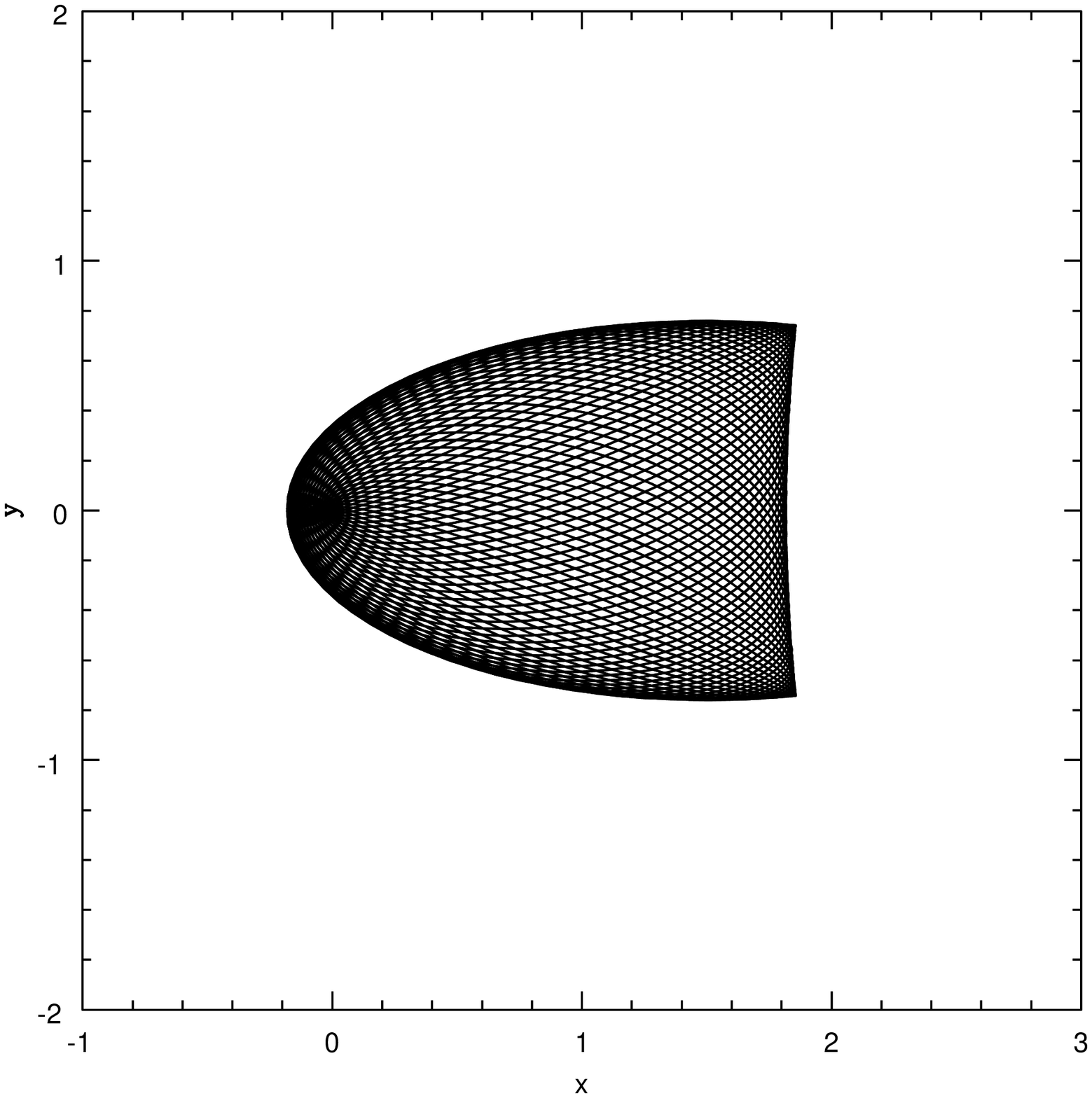}}}
\centerline{\hspace*{0.6in}$(a)$\hfill$(b)$\hspace{0.5in}}
\caption[Figure 4]{Lens orbits for the planar, lopsided, harmonic perturbation: (a) an
anti--aligned  lens orbit for $a=1$, $d=0.5$, (b) an aligned  lens orbit for $a=1$, $d=1.5$.}
\end{figure}

\subsection{Possible applications to lopsided galactic nuclei}

The aligned loops (which are fixed points of the slow dynamics) have
constant angular momentum; $\ell$ is positive (negative) for
anti-clockwise (clockwise) motion of the particle on its Keplerian
ellipse.  Their eccentricity, $e=d/a\,$, which means that the centres
of the ellipses coincides with the centre of the perturbation (the BH
is at one of the foci).  For a fixed value of $d$, aligned loops of
varying sizes (i.e. $a$), form a family of concentric, confocal
ellipses. Not only can stars be placed on such orbits, but being a
non--intersecting family, such nested ellipses can also be the
streamlines of gas flows. Figure~5a shows one such representative set
of aligned loops. Stable oscillations about these fixed point orbits
allows $\ell$ to oscillate, while preserving a definite $\pm$ sign.
Unlike the circulating loops of the centred perturbation (see
Figure~1c), $g$ for one of these {\em librating loops} oscillates
about its mean value of $\pi$ (see Figure~5b).

\begin{figure}
\centerline{\hbox{\epsfxsize=1.7in\epsfbox{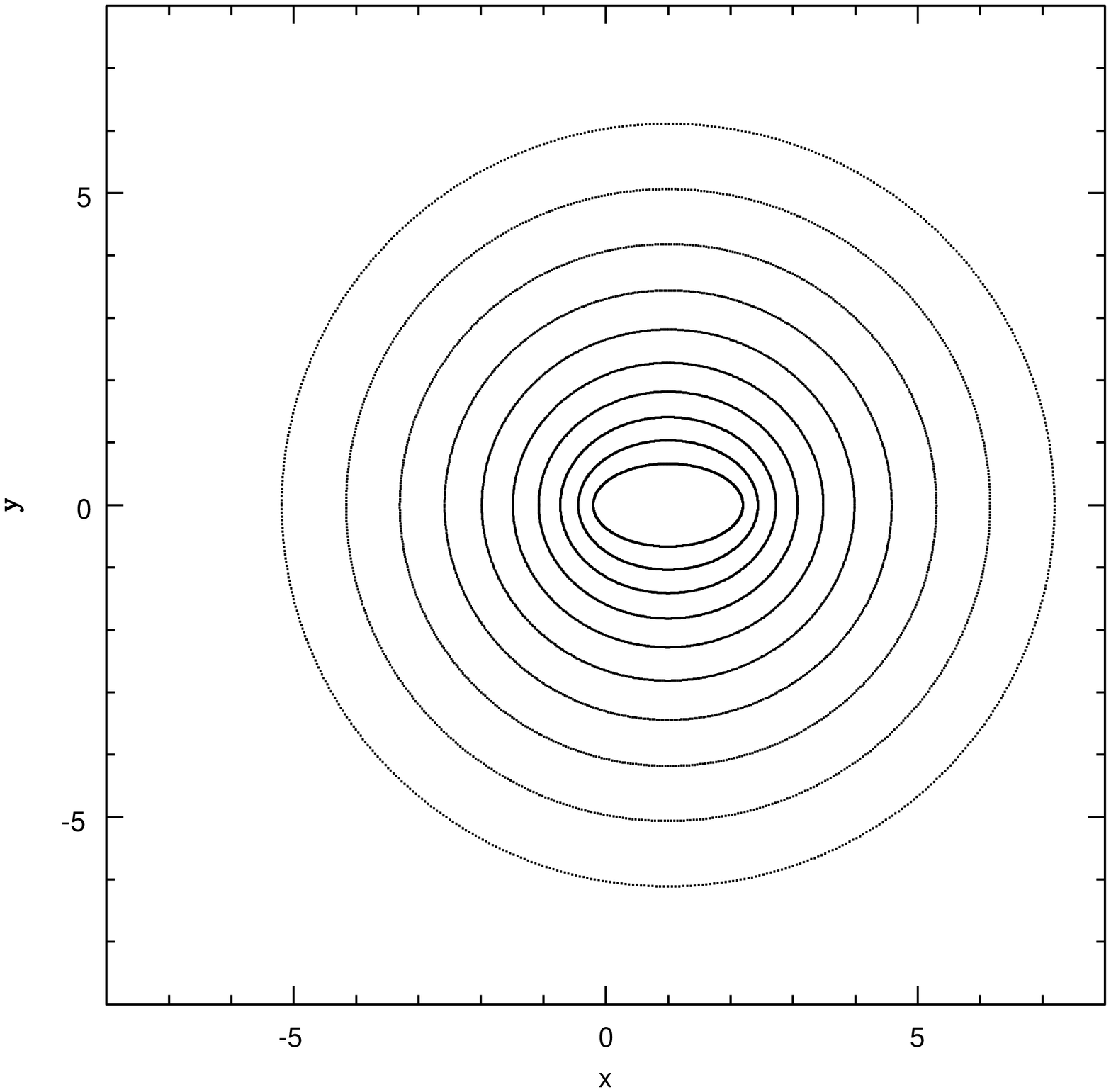}
                  \epsfxsize=1.7in\epsfbox{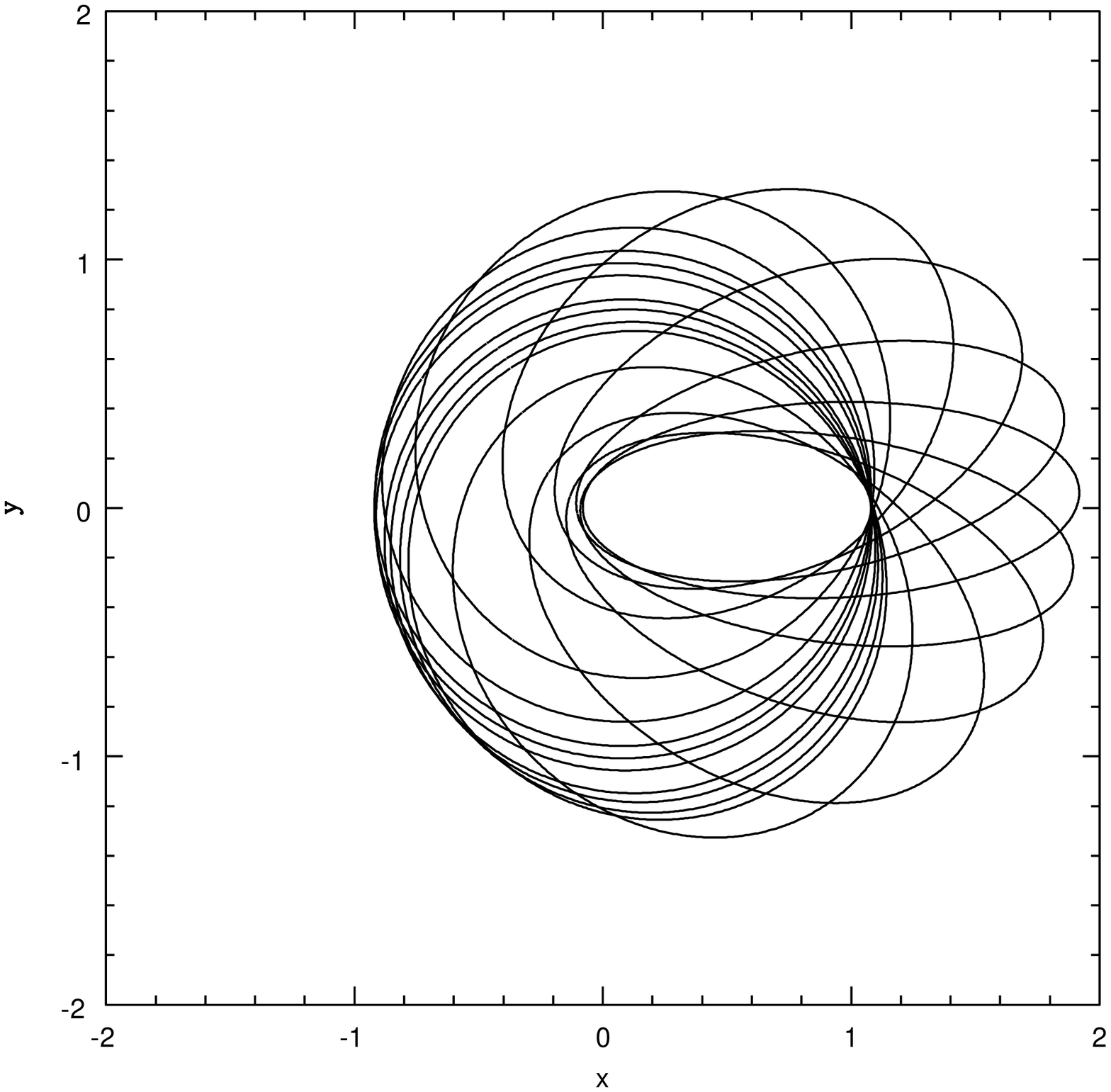}}}
\centerline{\hspace*{0.6in}$(a)$\hfill$(b)$\hspace{0.5in}}
\caption[Figure 5]{Loop orbits for the planar, lopsided, harmonic 
perturbation: (a) a set of  aligned--loops when $d=1$. The ten nested 
ellipses have $a= 1.2, (1.2)^2, \ldots , (1.2)^{10}\simeq 6.19$, 
(b) A librating loop for $d=0.5$, $a=1$, for which $\ell$ fluctuates 
between $0.4$ and $0.97$.} 
\end{figure}

Tremaine~(1995) has suggested that the nucleus of M31 is an eccentric
disc in which stars move on nearly Keplerian orbits around a massive
BH.  In his model, the location of the BH coincides with the fainter
(brightness) peak, P2, and brighter peak, P1, is the apoapsis region
of the Keplerian ellipses. 
If we imagine that the
lopsided harmonic perturbation mocks self--gravity, our toy model 
suggests that the orbits such as our aligned loops (and their progeny,
the librating loops) might form the backbone of star clusters such
as Tremaine's eccentric discs.

\section{Centred perturbations for general $\alpha$}

When $\alpha\neq 2$, it is no longer possible to obtain exact expressions
for the slow Hamiltonian. One possibility is to evaluate $H$ numerically,
and study the contour plots. Another option is to look at limiting cases,
which is what we do below. In this Section we study centred,
nearly circular perturbations: $d_1=d_2=0$ and $0<\epsilon\ll 1$. 
Equation~(\ref{potential.2}), for
$\alpha\neq 0$, gives us the following scale--free potentials:
\begin{eqnarray}
\frac{\Phi(x,y)}{\mu I} &=& \frac{1}{a^{\alpha}}\left(x^2 + (1+\epsilon)y^2
\right)^{\alpha/2} = \frac{1}{a^{\alpha}}\left(r^2 + \epsilon y^2
\right)^{\alpha/2} \nonumber \\[1ex]
{} &=&  \left(\frac{r}{a}\right)^{\alpha} \;+\; \epsilon\frac{\alpha}{2}
\left(\frac{r}{a}\right)^{\alpha -2}\left(\frac{y}{a}\right)^2 \;+\;
O(\epsilon^2)\,.\label{smal.eps}
\end{eqnarray}
\noindent We can now substitute in equation~(\ref{smal.eps}), the following expressions,
\begin{eqnarray}
r &=& a\left(1-\sqrt{1-\ell^2}\cos\eta\right)\,,\nonumber \\
y &=& a\left\{{\sin g}\,\left(\cos\eta - \sqrt{1-\ell^2}\right) +\ell{\cos g}\,\sin\eta\right\}\,,
\label{ry2del}
\end{eqnarray}
\noindent  for $r$ and $y$, and average over $\eta$ 
to obtain the averaged Hamiltonian (see Appendix~A2 for 
the derivation) correct to first order in $\epsilon$:
\begin{eqnarray}
H(\ell, g;\alpha, \epsilon) &=& \left(1+\frac{\epsilon\alpha}{4}
-\frac{\epsilon\alpha}{4}\,\cos 2 g\right)\;\times\nonumber \\[1ex]
&\times&\; 
F\left(\;-\,\frac{1+\alpha}{2}\;, \;-\frac{\alpha}{2}\;, \;1\;, \;1-\ell^2\,\right)
\nonumber \\[1ex] 
&+&\frac{\epsilon\alpha}{4}\,\ell^2\cos 2 g\,\times\nonumber \\[1ex]
&\times &\; F\left(\;\frac{1-\alpha}{2}\;, \;\frac{2-\alpha}{2}\;, \;2\;, \;1-\ell^2\;\,\right)
\,,\label{hsmal.eps}
\end{eqnarray}
\noindent where $F$ is Gauss' Hypergeometric function. When $\alpha=1$
or $2$, the Hypergeometric series for $F$ terminates, and the
Hamiltonian assumes a simple form. We have already discussed the case
$\alpha=2$ in some detail.\footnote{For $\alpha=2$,
equation~(\ref{hsmal.eps}) reduces to equation~(\ref{cent.harmonic}),
which happens to be an exact expression, because terms of
$O(\epsilon^2)$ are identically zero for this case.}  When $\alpha=1$,
\begin{eqnarray}
F\left(-1, \;-1/2\;, \;1\;, \;1-\ell^2\,\right)
&=&\frac{3-\ell^2}{2}\,,\nonumber \\[1ex]
F\left(0, \;1/2\;, \;2\;, \;\;1-\ell^2\,\right) &=& 1
\,,\label{hyper.1}
\end{eqnarray}
\noindent so that
\begin{equation}
H(\ell, g; 1, \epsilon) \;=\; \left(\half+\frac{\epsilon}{8}\right)
\left(3-\ell^2\right) \;-\; \frac{3\epsilon}{8}\left(1-\ell^2\right)\,\cos 2g \,,
\label{ham.1}
\end{equation}

\begin{figure}
\centerline{\hbox{\epsfxsize=1.7in\epsfbox{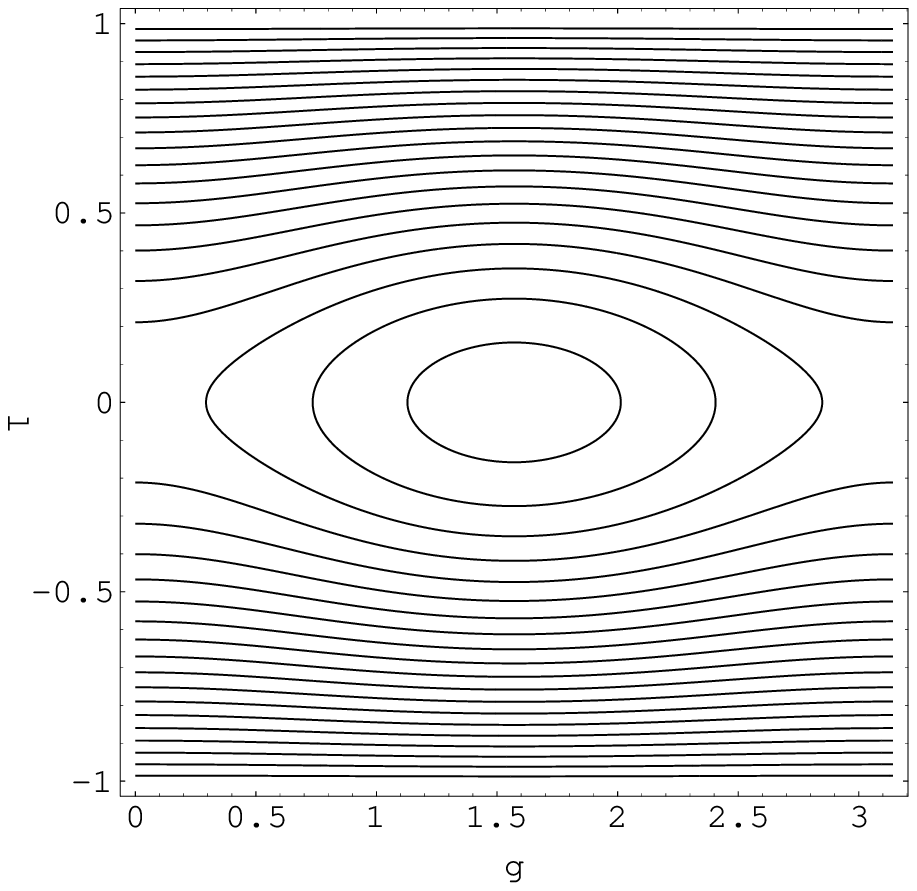}
                  \epsfxsize=1.7in\epsfbox{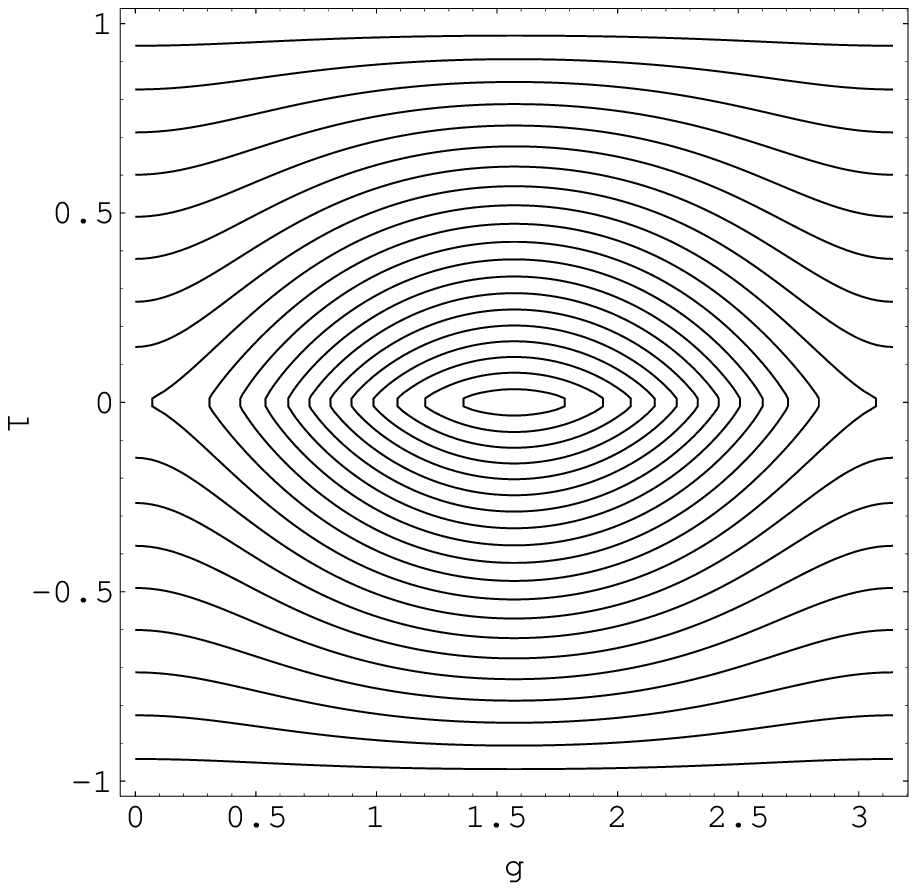}}}
\centerline{\hspace*{0.6in}$(a)$\hfill$(b)$\hspace{0.5in}}
\caption[Figure 6]{Isocontours of $H$ for $\epsilon=0.1$ (accurate to 
first order in $\epsilon$): (a) $\alpha=1$, (b) $\alpha=-0.9$.}
\end{figure}

\noindent to first order in $\epsilon$.
Figure~6a shows the isocontours of this Hamiltonian ($\alpha=1$) for
$\epsilon=0.1\,$; as may be seen, these are similar to those of
Figure~1a. Keeping $\epsilon$ small, we can vary $\alpha$ from $-1$ to
$2$. Figure~6b shows the isocontours for $\alpha=-0.9$; again the
overall structure is unchanged. Hence we conclude that, for small
$\epsilon$, the orbital structure is independent of $\alpha$ in the
range $(-1,2)$, with loops and short--axis lenses broadly similar to
those in Figures~1b and 1c.

\section{The limits of averaging}

The averaged Hamiltonian is an accurate proxy for the full Hamiltonian
when the orbital period is much shorter than the period of precession
of periapse.  Such is mostly the case for orbits that reside within the
sphere of influence of the BH. However as one moves outwards, the
influence of the perturbations relative to the BH's pull increases,
the mismatch in frequencies decreases, thereby invalidating the
averaging procedure. In practice, the breakdown occurs at resonances
between orbital and precessional motion and is responsible for the
emergence of minor orbit families such as the ones discussed by
Gerhard and Binney (1985) and Miralda-Escude and Schwarzschild
(1989)---hereafter GB and MES, respectively.

\begin{figure}
\centerline{\hbox{\epsfxsize=1.7in\epsfbox{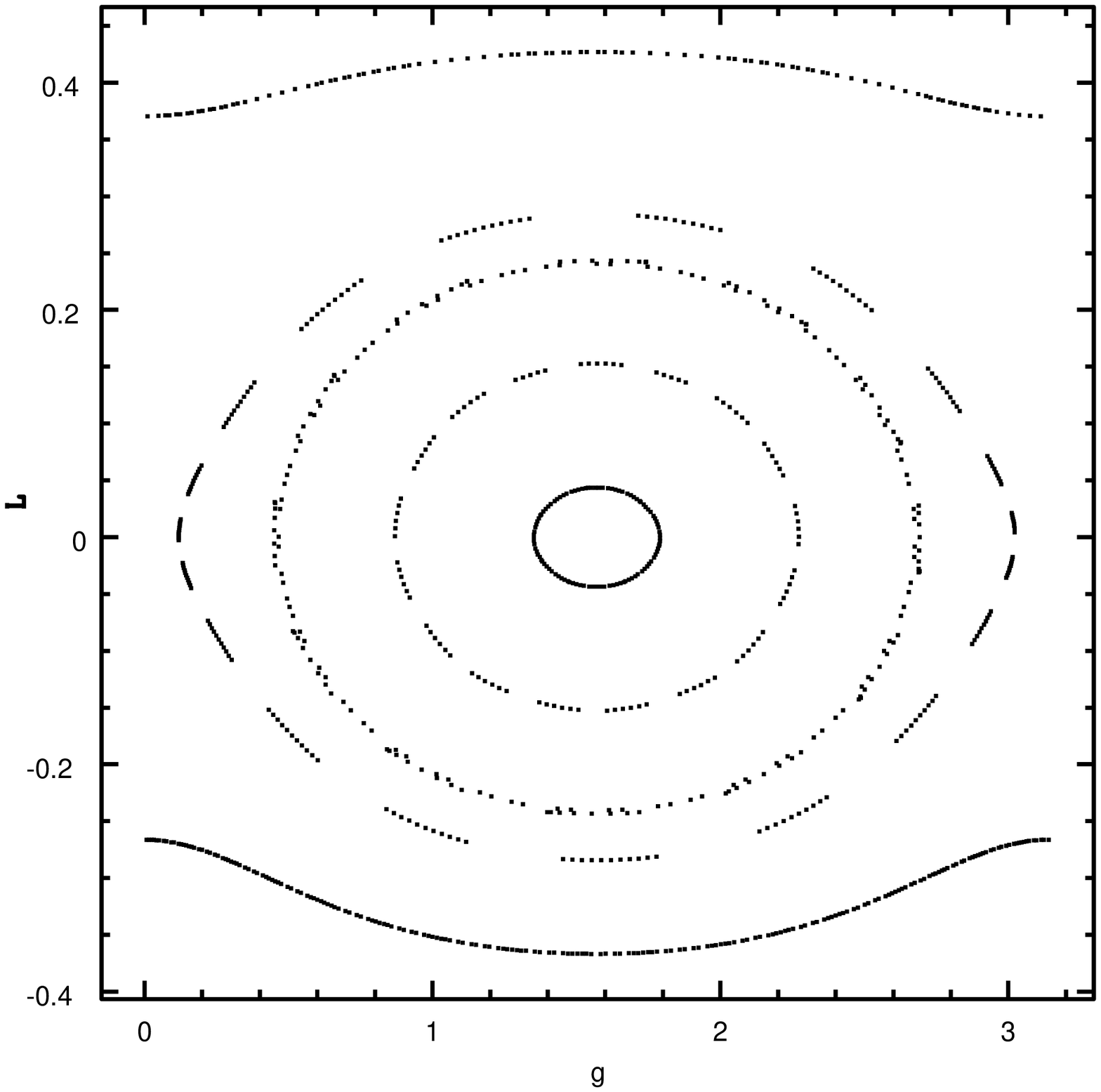}
                  \epsfxsize=1.7in\epsfbox{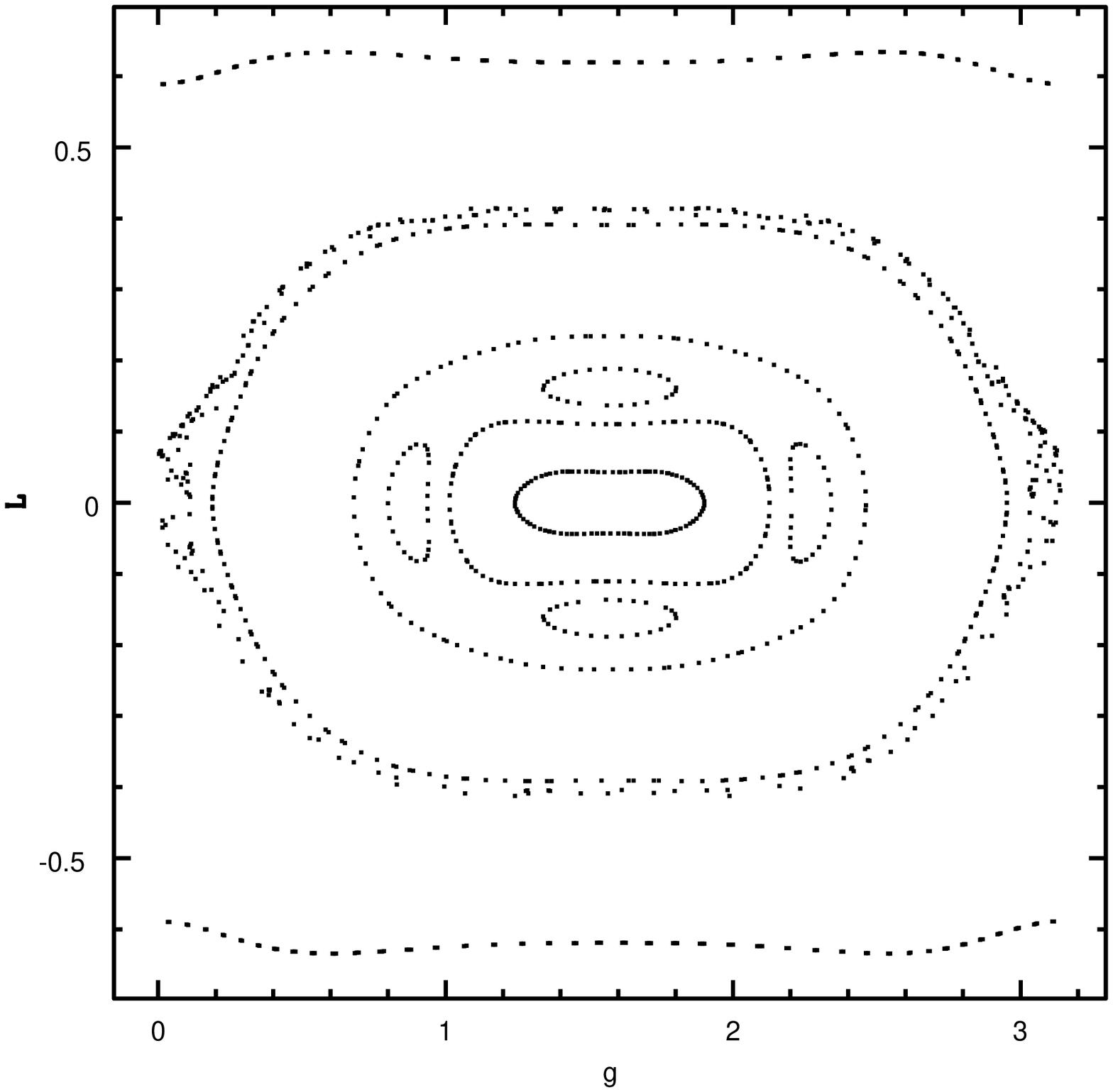}}}
\centerline{\hspace*{0.6in}$(a)$\hfill$(b)$\hspace{0.5in}}
\centerline{\hbox{\epsfxsize=1.7in\epsfbox{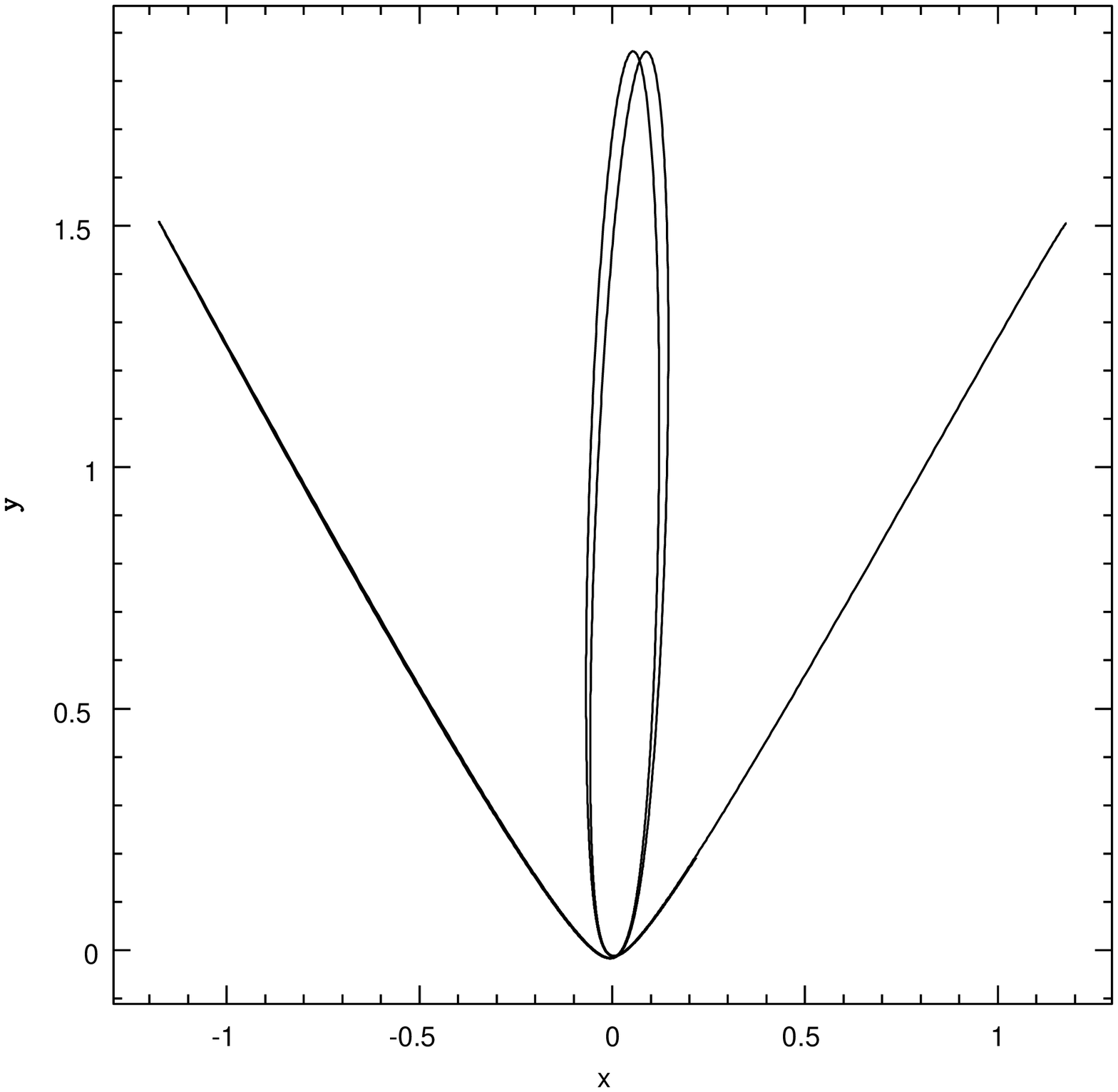}
                  \epsfxsize=1.7in\epsfbox{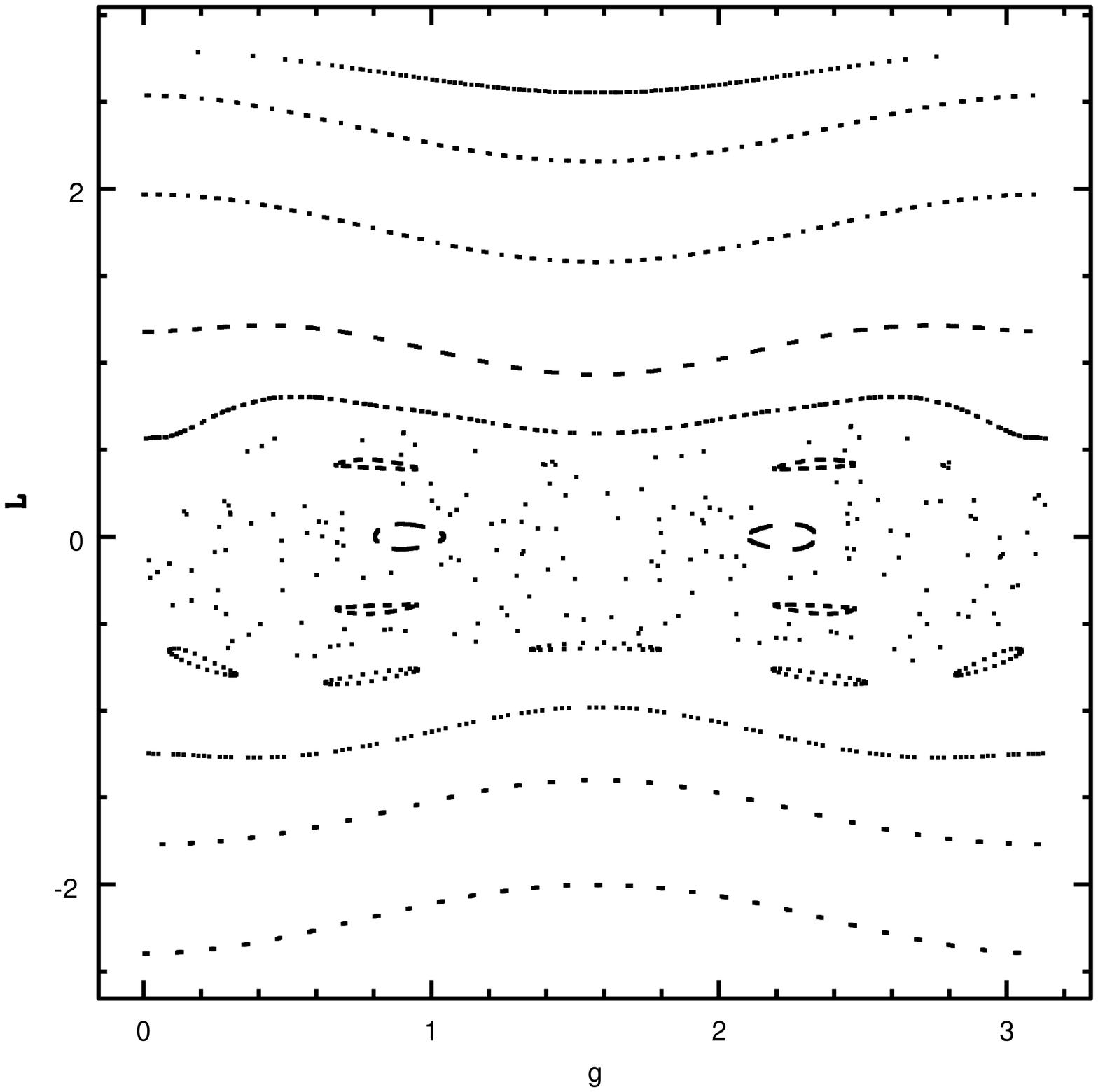}}}
\centerline{\hspace*{0.6in}$(c)$\hfill$(d)$\hspace{0.5in}}
\caption[Figure 7]{Dynamics in the potential of a black hole
perturbed by a logarithmic potential with $b=0.9$: (a) Section
on energy surface with zero velocity orbit $y=0$ and $x = 0.8r_h$; (b)
similar section but with $x = 2.0 r_h$ ; (c) Periodic orbit at the centre
of the chain of islands seen in (b); (d) Section on energy surface
with zero velocity orbit $y=0$ and $x = 4.0 r_h$.}
\end{figure}

In what follows, we display the gradual breakdown of averaging as one
moves away from a BH of mass M, in the scale--free logarithmic
potential of MES:
\begin{equation}
\Phi (x, y) =  \frac{v_L^2}{2}\ln\left(x^2 + \frac{y^2}{b^2}\right),
\end{equation}
\noindent where $v_L$ is the characteristic velocity of large loops, and $b$ is
a measure of non-axisymmetry. The dynamics being two dimensional, we
can explore the phase space structure on a Poincare section. Since we
are interested in resonances between orbital and precessional motions,
we choose to strobe the dynamics with the orbital period at each
apocentre passage, recording the angular momentum, $L$, and the
argument of the pericentre, $g$.

MES studied the scale--free case throroughly. The phase space is
divided between orbits with a definite sense of circulation (librating
non-aligned loops and rosettes) and those with no definite sense of
circulation. The latter are broken into resonant orbit families:
banana, fish, pretzel...etc. The resonances are between the orbital
period of the star and the period of precession of its pericentre.
The banana for instance, performs two orbital revolutions in the time
it takes its pericentre to precess once. Such resonances are naturally
missed by the averaged Hamiltonian which assumes that the orbital
period is much faster than the precessional period (see Touma and
Tremaine~(1997) for a discussion of the averaged Hamiltonian).

We now place a BH at the origin and study the restructing of orbits that
occurs. MES did introduce a BH but did not show any surface of
section. They presented samples of high energy orbits, including some
regular ones. The BH radius of influence is given by, $r_h=
GM/v_L^2$. We measure distance in units of $r_h$, and look at sections
of increasing energy. The energy is that of an orbit with zero
velocity started on the $x$-axis, with $x=f r_h$, $f=0.4, 2.0, 4.0$.

Within $r_h$ (Figures~7a), the phase space is occupied by two main
families of regular orbits: short axis lenses and loops. The lenses
appear naturally in the integrable cusps with BHs presented in Sridhar
and Touma~(1997). The apparent integrability of the motion is
consistent with the validity of averaging at distances where the
motion is dominated by the BH. The phase space of the corresponding
averaged Hamiltonian was not described in this paper, but it hardly
differs from the averaged dynamics of centered harmonic perturbations
discussed above (this should be obvious from the results of \S~4). As
one moves out in radius past $r_h$ (Figure~7b), resonances become
stronger, and a layer of stochastic motion develops near the
separatrix. A torus has broken up into a chain of islands. The
corresponding resonant periodic orbit, the bowtie, is shown in
Figure~7c. Further away from the BH (Figure~7d), the short axis radial
orbit becomes unstable, bananas are born, together with other resonant
orbits families which overlap and lead to large scale chaos where once
the regular lens orbits lived.

GB studied the logarithmic potential with core and BH. They were
mostly interested in energetic orbits far outside $r_h$ and they find
that resonant orbit families and stochastic orbits dominate the phase
space. When commenting on dynamics inside $r_h$, they just mention
that most orbits are loops which, as we now know, is incomplete. In
fact, within $r_h$, the dynamics is accurately described by the integrable,
orbit averaged Hamiltonian; the phase space is regular with mainly
short axis lenses and loops. As one moves outside the sphere of
influence, lenses and loops break down into resonant orbit families,
which get stronger, overlap and lead to stochastic orbits. The short
axis radial orbit goes unstable and regular lens orbits disappear, as
the region within the separatrix is practically engulfed by a single
stochastic orbit. GB described stochasticity in terms of scattering of
box orbits by the central cusp and BH. Here, stochasticity emerges
with the gradual sacrifice of regular lens orbits to strong
overlaping resonances as we move away from the BH.

\section{Discussion}

In this paper we have introduced an averaging technique into the study
of star clusters around massive BHs at the centres of galaxies. The
dynamics of these clusters is governed by the combined gravitational
potential of the central BH and the cluster itself. Within the sphere
of influence ($r_h$) of the BH, the cluster potential is a small, yet
non negligible perturbation. Orbits that are confined within $r_h$ may
be viewed as slowly precessing (and deforming) Keplerian ellipses. We
have shown that these orbits possess a quasi--integral of motion, in
addition to the energy integral; this nearly conserved quantity is
$I=\sqrt{GMa}$, where $M$ is the mass of the BH, and $a$ is the
semimajor axis of the orbit (when $M$ is constant, the semimajor axis
is nearly conserved). The quasi--integral arises because the Keplerian
orbital frequency is much larger than the frequencies of precessional
motions.  Taking advantage of this mismatch in frequencies, we
averaged over an orbital period, to obtain a reduced dynamics of slow,
precessional motions, wherein $I$ appears as a full--fledged integral
of motion.

The building of self--consistent stellar models for systems with low 
spatial symmetry (such as the central regions of M31 and NGC~4486B, which 
are known to possess double nuclei) surely requires 
the existence of more than one integral of motion. 
We now have in hand two integrals of motion---the energy and the semimajor
axis---that may be used to construct dynamical models for the centres of 
galaxies like  M31 and NGC~4486B. Let us verify that the conditions 
required for the averaging to apply are met in these two cases.

\noindent
1. M31 probably has a central BH of mass $\sim 6\times 10^7\msun$. For
velocity dispersion $\sim 160\kms$, $r_h\sim
10\pc$ ($\sim 3''$). The double nuclei are separated by $\sim 0.''5$
arcsec, and much of the nuclear disc itself appears to be within
$r_h$. Furthermore, within $1''$, the mass of the disc and the bulge
are only $16$~\%, and $2$~\% of the mass of the BH respectively (see
Tremaine~1995). Within $r_h$, the gravitational potentials of the disc
and the bulge are small, but non negligible, perturbations to the
potential of the BH; indeed, this is a situation that is ripe for the
application of the averaging technique! In \S~3.2 and 3.3, we
considered orbits in a toy model of a lopsided perturbation, and
discovered that there are a very interesting class of loop orbits that
are elongated in the same sense as the perturbation. The existence of
these aligned loop orbits is an encouraging sign that Tremaine's
(1995) eccentric disc can be realised as a self--consistent model.

\noindent
2. NGC 4486B (a low luminosity E1 companion of M87) also has a double
nucleus separated by $0.''15$ (Lauer et.al. 1996). There seems to be
some evidence for a central BH of mass $\sim 9\times 10^8\msun$
(Kormendy et. al. 1997).  For a velocity dispersion
$\sim 175\kms$, and an assumed distance to NGC~4486B of $16\mpc$, we
estimate that $r_h\sim 1.''6$, a length scale that is not only larger
than the separation between the nuclear peaks, but also one that is
nearly fifteen times larger than than the best resolution obtainable with
the HST; hence the averaging technique may be expected to be useful
here too.  Kormendy et. al. (1997) suggest that the double nucleus
might arise from a stellar distribution similar to Tremaine's (1995)
disc.  However, as Lauer et.al. (1996) note, there are differences
from the case of M31, and even the evidence for a massive BH is less
certain.

{\em A fundamental result of this paper is that 
the presence of a massive BH enforces a certain degree of 
integrability within its sphere of influence; and this happens because of
the existence of the secular invariant, $I$}. Even without detailed 
knowledge of the  (perturbing cluster) potential, 
we were able to note that slow dynamics in  (i) razor--thin discs is fully
integrable because the problem is two dimensional, (ii) any axisymmetric
potential is fully integrable (note that axisymmetry allows for 
lopsidedness), and (iii) a potential with no spatial symmetry whatsoever 
still conserves two integrals, so that any precessional chaos must be 
highly limited in nature. 

When the potential is time dependent, but with variations occurring over
times longer than the orbital times, averaging is still applicable. We
discussed the case of the adiabatic growth of the BH. Another case of
slow time variation comes to mind: this is the `resonant relaxation'
of Rauch and Tremaine (1996), a process that relies on two body
encounters between stars.  The time scales are typically much longer
than orbital times, so averaging applies, and the stars in their
courses may be treated as elliptical rings. Over certain time scales
(shorter than the classical two body relaxation times), the semimajor
axes of all the stars are approximately constant, whereas torques
between the rings exchange angular momentum. As the authors note, this
can lead to a situation in which some regions of nuclear star clusters
are relaxed in angular momentum, but not in energy.

We introduced a family of model planar potentials, that allow
for a range of cusp slopes ($\alpha$), non--axisymmetry and
lopsidedness, with a view to classification of orbit families. For the
harmonic case ($\alpha=2$), we showed that centered non--axisymmetric
perturbations support two main families of orbits: short axis lenses
and loops. Lenses were introduced in Sridhar and Touma (1997)
and should be recognized as an essential feature of regular motion
around BHs. They have zero average angular momentum and fill out lens
shaped regions in configuration space. The loops are the familiar
rosettes of slightly perturbed Keplerian motion. We explored the
consequences of lopsidedness and identified a family of
(resonant) aligned loops whose elongation is in the same sense as the
lopsidedness. 

When $\alpha\neq 2$, the case of small non--axisymmetry ($\epsilon\ll
1$) is analytically tractable. The
qualitative nature of the orbits (lenses and loops) appears to be
insensitive to $\alpha$.  This is probably because the force exerted
on a star due to the perturbing potential---even when it becomes
infinitely large at the centre---is overwhelmed by the force due to
the BH. If such is the case for larger non--axisymmetry, then our
explorations of the harmonic case provide a sketch of the possible
orbit families in razor--thin discs around massive BHs.  Whether the
qualitative nature of dynamics around BHs is indeed independent of the
steepness of the cusp in density is an unsolved problem, but one that
is tractable by methods such as our averaging technique.

Averaging is of course valid as long as we are far away
from resonances between the orbital period and the
periods of apsidal or nodal precession. We showed, numerically, 
that such a condition is indeed satisfied for  $r< r_h$, but  breaks down 
outside. Orbits which take a star far outside $r_h$ may be expected to
be chaotic (c.f. \cite{gerbin85}, \cite{mer97}).
Central BHs and strong cusps might force the main body on an elliptical
galaxy to evolve toward axisymmetry. However, as we have argued in this 
paper, the central regions of galaxies might
stubbornly persist with their striking variety of non--axisymmetric features.

\section{ACKNOWLEDGMENTS}

We thank the Raman Research Institute, and the Indian Railways for
their hospitality while this work was in progress. JT wishes to thank
IUCAA for support and hospitality at Pune, as well as acknowledge the
support of the Harlan Smith Fellowship under NASA grant NAGW 1477.

\appendix

\section{The case $\epsilon=3$}

We show that even the {\em unaveraged} dynamics, in the combined potential of a
BH and a centered harmonic perturbation, is integrable, when
$\epsilon=3$ (i.e. $b = 0.5$). Indeed the combined potential,
\begin{equation}
\Phi = -\frac{GM}{\sqrt{x^2+y^2}} + K(x^2 + 4 y^2), 
\end{equation}
when expressed in parabolic coordinates (see Sridhar and Touma 1997
for details), $\xi = y + \sqrt{x^2+y^2}$, and $\eta = y -
\sqrt{x^2+y^2}$, takes the form
\begin{equation} 
\Phi= \frac{F_{+}(\xi)}{\xi - \eta} + \frac{F_{-}(\eta)}{\eta - \xi}\,,
\end{equation}
where
\begin{eqnarray}
F_{+}(\xi) &=& K(\xi^{3}-GM)\,\nonumber \\
F_{-}(\eta) &=& K(-|\eta|^{3}+GM)\,.
\end{eqnarray}
This form is that of a separable, St\"ackel potential in parabolic
coordinates, which supports integrable motion. 

\section{Hamiltonian for general $\alpha$}

Let $\Phi$ denote the potential given in equation~(\ref{smal.eps}).
Substitute for $r$ and $y$, the expressions given in
equations~(\ref{ry2del}), and average over $\eta$ in the usual manner.
This gives us the averaged Hamiltonian, $H=(\overline{\Phi}/\mu I)=
H_0 +\epsilon H_1 + O(\epsilon^2)\,$, where
\begin{eqnarray}
H_0 &=& \frac{1}{2\pi}\oint d\eta\left(1-\sqrt{1-\ell^2}\cos\eta
\right)^{\alpha +1}\,,\nonumber \\[1em]
H_1 &=& \frac{\alpha}{4\pi}\oint d\eta\left(1-\sqrt{1-\ell^2}\cos\eta
\right)^{\alpha -1}\times\nonumber\\[1ex]
\quad&\times&\left\{{\sin g}\,\left(\cos\eta - \sqrt{1-\ell^2}\right) 
+\ell\,\cos g\,\right\}^2\,.\label{eqn.h01}
\end{eqnarray}
\noindent Some straightforward algebra allows us to express
\begin{equation}
H = \left(1+\frac{\epsilon\alpha}{4}\right)A(\ell; \alpha) -
\frac{\epsilon\alpha}{4} B(\ell; \alpha)\cos 2 g
\;+\; O(\epsilon^2)\,,
\label{eqnhnew}
\end{equation}
\noindent in terms of the two functions, $A$ and $B$:
\begin{eqnarray}
A(\ell; \alpha) &=&  F\left(\;-\,\frac{1+\alpha}{2}\;, 
\;-\frac{\alpha}{2}\;, \;1\;, \;1-\ell^2\,\right)\,,\nonumber \\[1ex]
B(\ell; \alpha) &=& A \;-\; \ell^2\;
F\left(\frac{1-\alpha}{2}, \frac{2-\alpha}{2}, 2,\;1-\ell^2\right)\,,
\label{ab.final}
\end{eqnarray}
\noindent where we have used formulae 3.666, 8.703 and 9.131~(1) of 
Gradshteyn and Ryzhik~(1994). Substituting for $A$ and $B$ in 
equation~(\ref{eqnhnew})
completes the derivation.


\begin{thebibliography}{}


\bibitem[Gebhardt et al. (1996)]{geb96}Gebhardt, K., et. al., 1996, AJ, 112, 105

\bibitem[Gerhard \& Binney (1985)]{gerbin85}Gerhard, O. E., \& Binney,
J., 1985, MNRAS, 216, 467

\bibitem[Goldstein (1980)]{gold80}Goldstein, H., 1980, Classical Mechanics, 
2nd ed. (Reading: Addison-Wesley)

\bibitem[]{}Gradshteyn, I. S., and Ryzhik, I. M., 1994, Table of Integrals,
Series, and Products, edited by A. Jeffrey (Academic Press, Inc, Boston)

\bibitem[Hagihara (1971)]{HAGIHARA} Hagihara, Y., 1971, Celestial Mechanics II, Part I
(MIT Press, Cambridge, Ma)

\bibitem[]{}Heisler, J. and Tremaine, S., 1986, Icarus, 65, 13

\bibitem[]{}Kormendy,~J., 1982, in Morphology and Dynamics of Galaxies,
edited by L.~Martinet and M.~Mayor (Geneva Observatory, Sauverny), p 113

\bibitem[]{}Kormendy,~J., 1987, in Structure and Dynamics of Elliptical
Galaxies, IAU Symposium No. 127, edited by T. de Zeeuw (Reidel, Dordrecht), 
p 17

\bibitem[Kormendy \& Richstone (1995)]{korric95}Kormendy, J., \& Richstone,
D., 1995, ARAA 33, 581

\bibitem[]{}Kormendy,~J., et. al., 1997, ApJ, 482, L139 

\bibitem[]{}Lauer,~T.R., 1983, Ph.D thesis, University of California, Santa Cruz

\bibitem[]{}Lauer,~T.R., 1985, ApJ, 292, 104

\bibitem[]{}Lauer,~T.R. et. al. 1993, AJ, 106, 1436

\bibitem[Lauer et. al.  (1995)]{lauer95}Lauer,~T.R., et. al., 1995,
AJ, 110, 2622

\bibitem[Lauer et. al.  (1996)]{lauer96}Lauer,~T.R., et. al., 1996, ApJ, 471, L79


\bibitem[Merritt (1997)]{mer97}Merritt, D., 1997, ApJ, 486, 102

\bibitem[Merritt (1998)]{mer98}Merritt, D., 1998, Comments on Astrophysics, 
19

\bibitem[Miralda-Escud\'e \& Schwarzschild (1989)]{mirsch89}Miralda-Escud\'e,
J., \& Schwarzschild,~M., 1989, ApJ 339, 752

\bibitem[]{}Peebles,~P.J.E., 1972, Gen. Rel. Grav., 3, 61

\bibitem[]{}Pfenniger,~D. \& de Zeeuw, T., 1989, in Dynamics of Dense Stellar
Systems, edited by D. Merritt (Cambridge University Press, Cambridge), 81

\bibitem[Plummer (1960)]{plum60}Plummer, H. C., 1960, An Introductory 
Treatise on Dynamical Astronomy (Dover, New York)

\bibitem[]{}Rauch,~K.P., \& Tremaine,~S., 1996, New Astronomy, 1, 149

\bibitem[]{}Rees,~M.J., 1990, Science, 247, 817

\bibitem[Sridhar \& Touma (1997)]{sritou97} Sridhar, S., \& Touma, J., 1997, MNRAS 287, L1-L4

\bibitem[]{}Touma,~J. \& Tremaine,~S., 1997, MNRAS 292, 905

\bibitem[]{}Tremaine,~S., 1995, AJ, 110, 628

\bibitem[]{}Young,~P., 1980, ApJ, 242, 1232

\end{thebibliography}
\end{document}